\batchmode
\makeatletter
\def\input@path{{\string"/home/hari/Git/papers/A Fast Numerical Scheme for the Godunov-Peshkov-Romenski Model of Continuum Mechanics/\string"}}
\makeatother
\documentclass[twoside,english,3p,times]{elsarticle}
\usepackage[T1]{fontenc}
\usepackage[latin9]{inputenc}
\usepackage{prettyref}
\usepackage{amsmath}
\usepackage{amssymb}
\usepackage{graphicx}
\usepackage{wasysym}

\makeatletter

\providecommand{\tabularnewline}{\\}

\journal{Journal of Computational Physics}

\usepackage{fancyhdr}
\usepackage{xcolor}

\makeatother

\usepackage{babel}
\begin{document}

\begin{frontmatter}{}

\title{A Fast Numerical Scheme for the Godunov-Peshkov-Romenski Model of
Continuum Mechanics}

\author{Haran Jackson}

\ead{hj305@cam.ac.uk}

\address{Cavendish Laboratory, JJ Thomson Ave, Cambridge, UK, CB3 0HE}
\begin{abstract}
A new second-order numerical scheme based on an operator splitting
is proposed for the Godunov-Peshkov-Romenski model of continuum mechanics.
The homogeneous part of the system is solved with a finite volume
method based on a WENO reconstruction, and the temporal ODEs are solved
using some analytic results presented here. Whilst it is not possible
to attain arbitrary-order accuracy with this scheme (as with ADER-WENO
schemes used previously), the attainable order of accuracy is often
sufficient, and solutions are computationally cheap when compared
with other available schemes. The new scheme is compared with an ADER-WENO
scheme for various test cases, and a convergence study is undertaken
to demonstrate its order of accuracy.
\end{abstract}

\end{frontmatter}{}

\global\long\def\dev{\operatorname{dev}}

\global\long\def\tr{\operatorname{tr}}

\global\long\def\erf{\operatorname{erf}}

\tableofcontents{}

\rule[0.5ex]{1\columnwidth}{1pt}

\section{Background}

\subsection{Motivation}

The Godunov-Peshkov-Romenski model of continuum mechanics (as described
in \prettyref{subsec:The-GPR-Model}) presents an exciting possibility
of being able to describe both fluids and solids within the same mathematical
framework. This has the potential to streamline development of simulation
software by reducing the number of different systems of equations
that require solvers, and cutting down on the amount of theoretical
work required, for example in the treatment of interfaces in multimaterial
problems. In addition to this, the hyperbolic nature of the GPR model
ensures that the nonphysical instantaneous transmission of information
appearing in certain non-hyperbolic models (such as the Navier-Stokes
equations) cannot occur. Parallelization also tends to be easier with
hyperbolic models, allowing us to leverage the great advances that
have been made in parallel computing architectures in recent years.

At the time of writing, the GPR model has been solved for a variety
of fluid and solid problems using the ADER-WENO method (\citet{Dumbser2016a,Boscheri2016}).
ADER-WENO methods (described in \prettyref{subsec:The-ADER-WENO-Method})
are extremely effective in producing arbitrarily-high order solutions
to hyperbolic systems of PDEs, but in some situations their accompanying
computational cost may prove burdensome. A new method is presented
in this study that is simple to implement and computationally cheaper
than a corresponding ADER-WENO method if only second order accuracy
is required. This may prove useful in the design of simulation software
addressing problems in which not just accuracy but also speed and
usability are of paramount importance.

\subsection{The GPR Model\label{subsec:The-GPR-Model}}

The GPR model, first introduced in \citet{Peshkov2016}, has its roots
in Godunov and Romenski's 1970s model of elastoplastic deformation
(see \citet{godunov2003elements}). It was expanded upon in \citet{Dumbser2016a}
to include thermal conduction. This expanded model takes the following
form:

\begin{subequations}

\begin{align}
\frac{\partial\rho}{\partial t}+\frac{\partial\left(\rho v_{k}\right)}{\partial x_{k}} & =0\label{eq:DensityEquation}\\
\frac{\partial\left(\rho v_{i}\right)}{\partial t}+\frac{\partial(\rho v_{i}v_{k}+p\delta_{ik}-\sigma_{ik})}{\partial x_{k}} & =0\label{eq:MomentumEquation}\\
\frac{\partial A_{ij}}{\partial t}+\frac{\partial\left(A_{ik}v_{k}\right)}{\partial x_{j}}+v_{k}\left(\frac{\partial A_{ij}}{\partial x_{k}}-\frac{\partial A_{ik}}{\partial x_{j}}\right) & =-\frac{\psi_{ij}}{\theta_{1}(\tau_{1})}\label{eq:DistortionEquation}\\
\frac{\partial\left(\rho J_{i}\right)}{\partial t}+\frac{\partial\left(\rho J_{i}v_{k}+T\delta_{ik}\right)}{\partial x_{k}} & =-\frac{\rho H_{i}}{\theta_{2}\left(\tau_{2}\right)}\label{eq:ThermalEquation}\\
\frac{\partial\left(\rho E\right)}{\partial t}+\frac{\partial\left(\rho Ev_{k}+\left(p\delta_{ik}-\sigma_{ik}\right)v_{i}+q_{k}\right)}{\partial x_{k}} & =0\label{eq:EnergyEquation}
\end{align}

\end{subequations}

$\rho$,$\mathbf{v}$,$p$,$\delta$,$\sigma$,$T$,$E$,$\boldsymbol{q}$
retain their usual meanings. $\theta_{1}$ and $\theta_{2}$ are positive
scalar functions, chosen according to the properties of the material
being modeled. $A$ is the distortion tensor (containing information
about the deformation and rotation of material elements), $\mathbf{J}$
is the thermal impulse vector (a thermal analogue of momentum), $\tau_{1}$
is the strain dissipation time, and $\tau_{2}$ is the thermal impulse
relaxation time. $\psi=\frac{\partial E}{\partial A}$ and $\boldsymbol{H}=\frac{\partial E}{\partial\boldsymbol{J}}$.

The following definitions are given:

\begin{subequations}

\begin{align}
p & =\rho^{2}\frac{\partial E}{\partial\rho}\\
\sigma & =-\rho A^{T}\frac{\partial E}{\partial A}\\
T & =\frac{\partial E}{\partial s}\\
\boldsymbol{q} & =\frac{\partial E}{\partial s}\frac{\partial E}{\partial\boldsymbol{J}}
\end{align}

\end{subequations}

To close the system, the equation of state (EOS) must be specified,
from which the above quantities and the sources can be derived. $E$
is the sum of the contributions of the energies at the molecular scale
(microscale), the material element\footnote{The concept of a \textit{material element} corresponds to that of
a fluid parcel from fluid dynamics, applied to both fluids and solids.} scale (mesoscale), and the flow scale (macroscale):

\begin{equation}
E=E_{1}\left(\rho,p\right)+E_{2}\left(A,\boldsymbol{J}\right)+E_{3}\left(\boldsymbol{v}\right)\label{eq:EOS}
\end{equation}

The EOS used in this study (and described in the following passages)
is taken from \citet{Dumbser2016a}. It should be noted, however,
that this is just one particular choice, and there are many others
that may be used.

For an ideal or stiffened gas, $E_{1}$ is given by:

\begin{equation}
E_{1}=\frac{p+\gamma p_{\infty}}{\left(\gamma-1\right)\rho}
\end{equation}

where $p_{\infty}=0$ for an ideal gas.

$E_{2}$ is chosen to have the following quadratic form:

\begin{equation}
E_{2}=\frac{c_{s}^{2}}{4}\left\Vert \dev\left(G\right)\right\Vert _{F}^{2}+\frac{\alpha^{2}}{2}\left\Vert \boldsymbol{J}\right\Vert ^{2}
\end{equation}
$c_{s}$ is the characteristic velocity of propagation of transverse
perturbations. $\alpha$ is a constant related to the characteristic
velocity of propagation of heat waves:

\begin{equation}
c_{h}=\frac{\alpha}{\rho}\sqrt{\frac{T}{c_{v}}}
\end{equation}

$G=A^{T}A$ is the Gramian matrix of the distortion tensor, and $\dev\left(G\right)$
is the deviator (trace-free part) of $G$:

\begin{equation}
\dev\left(G\right)=G-\frac{1}{3}\tr\left(G\right)I
\end{equation}

$E_{3}$ is the usual specific kinetic energy per unit mass:

\begin{equation}
E_{3}=\frac{1}{2}\left\Vert \boldsymbol{v}\right\Vert ^{2}
\end{equation}

The following forms are chosen:

\begin{subequations}

\begin{align}
\theta_{1}\left(\tau_{1}\right) & =\frac{\tau_{1}c_{s}^{2}}{3\left|A\right|^{\frac{5}{3}}}\\
\theta_{2}\left(\tau_{2}\right) & =\tau_{2}\alpha^{2}\frac{\rho T_{0}}{\rho_{0}T}
\end{align}

\end{subequations}

\begin{subequations}

\begin{align}
\tau_{1} & =\frac{6\mu}{\rho_{0}c_{s}^{2}}\\
\tau_{2} & =\frac{\rho_{0}\kappa}{T_{0}\alpha^{2}}
\end{align}

\end{subequations}

The justification of these choices is that classical Navier\textendash Stokes\textendash Fourier
theory is recovered in the stiff limit $\tau_{1},\tau_{2}\rightarrow0$
(see \citet{Dumbser2016a}). This results in the following relations:

\begin{subequations}

\begin{align}
\sigma & =-\rho c_{s}^{2}G\dev\left(G\right)\\
\boldsymbol{q} & =\alpha^{2}T\boldsymbol{J}\\
-\frac{\psi}{\theta_{1}(\tau_{1})} & =-\frac{3}{\tau_{1}}\left|A\right|^{\frac{5}{3}}A\dev\left(G\right)\\
-\frac{\rho\boldsymbol{H}}{\theta_{2}\left(\tau_{2}\right)} & =-\frac{T\rho_{0}}{T_{0}\tau_{2}}\boldsymbol{J}
\end{align}

\end{subequations}

The following constraint also holds (see \citet{Peshkov2016}):

\begin{equation}
\det\left(A\right)=\frac{\rho}{\rho_{0}}
\end{equation}

The GPR model and Godunov and Romenski's 1970s model of elastoplastic
deformation in fact relie upon the same equations. The realization
of Peshkov and Romenski was that these are the equations of motion
for an arbitrary continuum - not just a solid - and so the model can
be applied to fluids too. Unlike in previous continuum models, material
elements have not only finite size, but also internal structure, encoded
in the distortion tensor.

The strain dissipation time $\tau_{1}$ of the HPR model is a continuous
analogue of Frenkel's ``particle settled life time'' \citet{jacovfrenkel1947};
the characteristic time taken for a particle to move by a distance
of the same order of magnitude as the particle's size. Thus, $\tau_{1}$
characterizes the time taken for a material element to rearrange with
its neighbors. $\tau_{1}=\infty$ for solids and $\tau_{1}=0$ for
inviscid fluids. It is in this way that the HPR model seeks to describe
all three major phases of matter, as long as a continuum description
is appropriate for the material at hand.

The evolution equation for $\boldsymbol{J}$ and its contribution
to the energy of the system are derived from Romenski's model of hyperbolic
heat transfer, originally proposed in \citet{Malyshev1984,Romenski1988},
and implemented in \citet{Romenski2007,Romenski2010}. In this model,
$\boldsymbol{J}$ is effectively defined as the variable conjugate
to the entropy flux, in the sense that the latter is the derivative
of the specific internal energy with respect to $\boldsymbol{J}$.
Romenski remarks that it is more convenient to evolve $\boldsymbol{J}$
and $E$ than the heat flux or the entropy flux, and thus the equations
take the form given here. $\tau_{2}$ characterizes the speed of relaxation
of the thermal impulse due to heat exchange between material elements.

\subsection{The ADER-WENO Method\label{subsec:The-ADER-WENO-Method}}

The ADER-WENO method was used in \citet{Dumbser2016a,Boscheri2016}
to solve the GPR system. It produces arbitrarily high-order solutions
to hyperbolic systems of PDEs and has been shown to be particularly
effective for a wide range of systems (e.g. the classical Euler equations
of gas dynamics, the special relativistic hydrodynamics and ideal
magnetohydrodynamics equations, and the Baer-Nunziato model for compressible
two-phase flow - see \citet{Balsara2009,Zanotti2016}). The first
step in the process - the WENO method - will be used later in this
study and is therefore discussed in detail here. The remaining steps
are described qualitatively, with references for further information
given.

WENO (Weighted Essentially Non-Oscillatory) methods are used to produce
high order polynomial approximations to piece-wise constant data.
Many variations exist. In this study, the method of \citet{Dumbser2013}
is used.

Consider the domain $\left[0,L\right]$. Take $K,N\in\mathbb{N}$.
The order of accuracy of the resulting method will be $N+1$. Take
the set of grid points $x_{i}=\frac{i\cdot L}{K}$ for $i=0,\ldots,K$
and let $\Delta x=\frac{L}{K}$. Denote cell $\left[x_{i},x_{i+1}\right]$
by $C_{i}$. Given cell-wise constant data $u$ on $\left[0,L\right]$,
an order $N$ polynomial reconstruction of $u$ in $C_{i}$ will be
performed. Define the scaled space variable:

\begin{equation}
\chi^{i}=\frac{1}{\Delta x}\left(x-x_{i}\right)
\end{equation}

Denoting the Gauss-Legendre abscissae on $\left[0,1\right]$ by $\left\{ \chi_{0},\ldots,\chi_{N}\right\} $,
define the nodal basis of order $N$: the Lagrange interpolating polynomials
$\left\{ \psi_{0},\ldots,\psi_{N}\right\} $ with the following property:

\begin{equation}
\psi_{i}\left(\chi_{j}\right)=\delta_{ij}
\end{equation}

If $N$ is even, take the stencils:

\begin{equation}
\begin{cases}
S_{1} & =\left\{ C_{i-\frac{N}{2}},\ldots,C_{i+\frac{N}{2}}\right\} \\
S_{2} & =\left\{ C_{i-N},\ldots,C_{i}\right\} \\
S_{3} & =\left\{ C_{i},\ldots,C_{i+N}\right\} 
\end{cases}
\end{equation}

If $N$ is odd, take the stencils:

\begin{equation}
\begin{cases}
S_{1} & =\left\{ C_{i-\left\lfloor \frac{N}{2}\right\rfloor },\ldots,C_{i+\left\lceil \frac{N}{2}\right\rceil }\right\} \\
S_{2} & =\left\{ C_{i-\left\lceil \frac{N}{2}\right\rceil },\ldots,C_{i+\left\lfloor \frac{N}{2}\right\rfloor }\right\} \\
S_{3} & =\left\{ C_{i-N},\ldots,C_{i}\right\} \\
S_{4} & =\left\{ C_{i},\ldots,C_{i+N}\right\} 
\end{cases}
\end{equation}

The data is reconstructed on $S_{j}$ as:

\begin{equation}
\sum_{p}\psi_{p}\left(\chi^{i}\left(x\right)\right)\hat{w}_{p}^{ij}
\end{equation}

where the $\hat{w}_{p}^{ij}$ are solutions to the following linear
system:

\begin{equation}
\frac{1}{\Delta x}\int_{x_{k}}^{x_{k+1}}\sum_{p}\psi_{p}\left(\chi^{k}\left(x\right)\right)\hat{w}_{p}^{ij}dx=u_{k}\qquad\forall C_{k}\in S_{j}
\end{equation}

where $u_{k}$ is the value of $u$ in $C_{k}$. This can be written
as $M_{j}\boldsymbol{\hat{w}^{ij}}=\boldsymbol{u_{\left[j_{0}:j_{N}\right]}}$
where $\left\{ j_{0},\ldots,j_{N}\right\} $ indexes the cells in
$S_{j}$. In this study reconstructions with $N=2$ are used. The
matrices of these linear systems are given in \prettyref{subsec:WENO-Matrices-for-N=00003D2},
along with their inverses, which are precomputed to accelerate the
solution of these systems.

Define the oscillation indicator matrix:

\begin{equation}
\Sigma_{mn}=\sum_{\alpha=1}^{N}\int_{0}^{1}\psi_{m}^{\left(\alpha\right)}\psi_{n}^{\left(\alpha\right)}d\chi
\end{equation}

and the oscillation indicator for each stencil:

\begin{equation}
o_{j}=\Sigma_{mn}\hat{w}_{m}^{ij}\hat{w}_{n}^{ij}
\end{equation}

The full reconstruction in $C_{i}$ is:

\begin{equation}
w_{i}\left(x\right)=\sum_{p}\psi_{p}\left(\chi^{i}\left(x\right)\right)\bar{w}_{p}^{i}
\end{equation}

where $\bar{w}_{p}^{i}=\omega_{j}\hat{w}_{p}^{ij}$ is the weighted
coefficient of the $p$th basis function, with weights:

\begin{equation}
\omega_{j}=\frac{\tilde{\omega}_{j}}{\sum_{k}\tilde{\omega}_{k}}\qquad\tilde{\omega}_{j}=\frac{\zeta_{j}}{\left(o_{j}+\varepsilon\right)^{r}}
\end{equation}

In this study, $r=8$, $\varepsilon=10^{-14}$, $\zeta_{j}=10^{5}$
if $S_{j}$ is a central stencil, and $\zeta_{j}=1$ if $S_{j}$ is
a side stencil, as in \citet{Dumbser2014}.

The reconstruction can be extended to two dimensions by taking:

\begin{equation}
\upsilon^{i}=\frac{1}{\Delta y}\left(y-y_{i}\right)
\end{equation}

and defining stencils in the y-axis in an analogous manner. The data
in $C_{i}$ is then reconstructed using stencil $S_{j}$ as:

\begin{equation}
\sum_{p,q}\psi_{p}\left(\chi^{i}\left(x\right)\right)\psi_{q}\left(\upsilon^{i}\left(x\right)\right)\tilde{w}_{pq}^{ij}
\end{equation}

where the coefficients of the weighted 1D reconstruction are used
as cell averages:

\begin{equation}
M_{j}\boldsymbol{\tilde{w}_{p}^{ij}}=\boldsymbol{\bar{w}_{p}^{\left[j_{0}:j_{N}\right]}}\qquad\forall p\in\left\{ 0,\ldots,N\right\} 
\end{equation}

The oscillation indicator is calculated for each $p$ in the same
manner as the 1D case. The reconstruction method is easily further
extensible to three dimensions, now using the coefficients $\bar{w}_{pq}$
of the weighted 2D reconstruction as cell averages.

The next process in the ADER-WENO method is to perform a Continuous
Galerkin or Discontinuous Galerkin spatio-temporal polynomial reconstruction
of the data in each cell, using the WENO reconstruction as initial
data at the start of the time step (see \citet{Balsara2009} and \citet{Dumbser2008}
respectively for implementations of these two variations). The order
of this reconstruction in time is usually taken to be the same as
the spatial order, and the same basis polynomials are used. The process
involves finding the root of a non-linear system, and this process
is guaranteed to converge in exact arithmetic for certain classes
of PDEs (see \citet{Jackson2017}). This root finding can be computationally
expensive relative to the WENO reconstruction, especially if the source
terms of the PDE system are stiff.

The final step in the ADER-WENO method is to perform a finite volume
update of the data in each cell, using the boundary-extrapolated values
of the cell-local Galerkin reconstructions to calculate the flux terms,
and the interior values of the Galerkin reconstructions to calculate
the interior volume integrals. See \citet{Dumbser2014} for more details.

\section{An Alternative Numerical Scheme}

Note that \eqref{eq:DensityEquation}, \eqref{eq:MomentumEquation},
\eqref{eq:DistortionEquation}, \eqref{eq:ThermalEquation}, \eqref{eq:EnergyEquation}
can be written in the following form:

\begin{equation}
\frac{\partial\boldsymbol{Q}}{\partial t}+\boldsymbol{\nabla}\cdot\boldsymbol{F}\left(\boldsymbol{Q}\right)+\boldsymbol{B}\left(\boldsymbol{Q}\right)\cdot\nabla\boldsymbol{Q}=\boldsymbol{S}\left(\boldsymbol{Q}\right)
\end{equation}

As described in \citet{Toro2009}, a viable way to solve inhomogeneous
systems of PDEs is to employ an operator splitting. That is, the following
subsystems are solved:

\begin{subequations}

\begin{align}
\frac{\partial\boldsymbol{Q}}{\partial t}+\boldsymbol{\nabla}\cdot\boldsymbol{F}\left(\boldsymbol{Q}\right)+\boldsymbol{B}\left(\boldsymbol{Q}\right)\cdot\nabla\boldsymbol{Q} & =\boldsymbol{0}\label{eq:HomogeneousSubsystem}\\
\frac{d\boldsymbol{Q}}{dt} & =\boldsymbol{S}\left(\boldsymbol{Q}\right)\label{eq:ODESubsystem}
\end{align}

\end{subequations}

The advantage of this approach is that specialized solvers can be
employed to compute the results of the different subsystems. Let $H^{\delta t},S^{\delta t}$
be the operators that take data $\boldsymbol{Q}\left(x,t\right)$
to $\boldsymbol{Q}\left(x,t+\delta t\right)$ under systems \eqref{eq:HomogeneousSubsystem}
and \eqref{eq:ODESubsystem} respectively. A second-order scheme (in
time) for solving the full set of PDEs over time step $\left[0,\Delta t\right]$
is obtained by calculating $\boldsymbol{Q_{\Delta t}}$ using a Strang
splitting:

\begin{equation}
\boldsymbol{Q_{\Delta t}}=S^{\frac{\Delta t}{2}}H^{\Delta t}S^{\frac{\Delta t}{2}}\boldsymbol{Q_{0}}
\end{equation}

In the scheme proposed here, the homogeneous subsystem will be solved
using a WENO reconstruction of the data, followed by a finite volume
update, and the temporal ODEs will be solved with appropriate ODE
solvers. This new scheme will be referred to here as \textit{the Split-WENO
method}.

\subsection{The Homogeneous System}

A WENO reconstruction of the cell-averaged data is performed at the
start of the time step (as described in \prettyref{subsec:The-ADER-WENO-Method}).
Focusing on a single cell $C_{i}$ at time $t_{n}$, we have $\boldsymbol{w^{n}}\left(\boldsymbol{x}\right)=\boldsymbol{w^{n}}_{p}\Psi_{p}\left(\boldsymbol{\chi}\left(\boldsymbol{x}\right)\right)$
in $C_{i}$ where $\Psi_{p}$ is a tensor product of basis functions
in each of the spatial dimensions. The flux in $C$ is approximated
by $\boldsymbol{F}\left(\boldsymbol{x}\right)\approx\boldsymbol{F}\left(\boldsymbol{w}_{p}\right)\Psi_{p}\left(\boldsymbol{\chi}\left(\boldsymbol{x}\right)\right)$.
$\boldsymbol{w}_{p}$ are stepped forwards half a time step using
the update formula:

\begin{equation}
\frac{\boldsymbol{w_{p}^{n+\frac{1}{2}}}-\boldsymbol{w_{p}^{n}}}{\Delta t/2}+\boldsymbol{F}\left(\boldsymbol{w_{k}^{n}}\right)\cdot\nabla\Psi_{k}\left(\boldsymbol{\chi_{p}}\right)+\boldsymbol{B}\left(\boldsymbol{w_{p}^{n}}\right)\cdot\left(\boldsymbol{w_{k}^{n}}\nabla\Psi_{k}\left(\boldsymbol{\chi_{p}}\right)\right)=\boldsymbol{0}
\end{equation}

i.e.

\begin{equation}
\boldsymbol{w_{p}^{n+\frac{1}{2}}}=\boldsymbol{w_{p}^{n}}-\frac{\Delta t}{2\Delta x}\left(\boldsymbol{F}\left(\boldsymbol{w_{k}^{n}}\right)\cdot\nabla\Psi_{k}\left(\boldsymbol{\chi_{p}}\right)+\boldsymbol{B}\left(\boldsymbol{w_{p}^{n}}\right)\cdot\left(\boldsymbol{w_{k}^{n}}\nabla\Psi_{k}\left(\boldsymbol{\chi_{p}}\right)\right)\right)\label{eq:WENO half step}
\end{equation}

where $\boldsymbol{\chi_{p}}$ is the node corresponding to $\Psi_{p}$.
This evolution to the middle of the time step is similar to that used
in the second-order MUSCL and SLIC schemes (see \citet{Toro2009})
and, as with those schemes, it is integral to giving the method presented
here its second-order accuracy. 

Integrating \eqref{eq:HomogeneousSubsystem} over $C$ gives:

\begin{equation}
\boldsymbol{Q_{i}^{n+1}}=\boldsymbol{Q_{i}^{n}}-\Delta t_{n}\left(\boldsymbol{P_{i}^{n+\frac{1}{2}}}+\boldsymbol{D_{i}^{n+\frac{1}{2}}}\right)
\end{equation}

where

\begin{subequations}

\begin{align}
\boldsymbol{Q_{i}^{n}} & =\frac{1}{V}\int_{C}\boldsymbol{Q}\left(\boldsymbol{x},t_{n}\right)d\boldsymbol{x}\\
\boldsymbol{P_{i}^{n+\frac{1}{2}}} & =\frac{1}{V}\int_{C}\boldsymbol{B}\left(\boldsymbol{Q}\left(\boldsymbol{x},t_{n+\frac{1}{2}}\right)\right)\cdot\nabla\boldsymbol{Q}\left(\boldsymbol{x},t_{n+\frac{1}{2}}\right)d\boldsymbol{x}\\
\boldsymbol{D_{i}^{n+\frac{1}{2}}} & =\frac{1}{V}\varoint_{\partial C}\boldsymbol{\mathcal{D}}\left(\boldsymbol{Q^{-}}\left(\boldsymbol{s},t_{n+\frac{1}{2}}\right),\boldsymbol{Q^{+}}\left(\boldsymbol{s},t_{n+\frac{1}{2}}\right)\right)d\boldsymbol{s}
\end{align}

\end{subequations}

where $V$ is the volume of $C$ and $\boldsymbol{Q^{-},Q^{+}}$ are
the interior and exterior extrapolated states at the boundary of $C$,
respectively.

Note that \eqref{eq:HomogeneousSubsystem} can be rewritten as:

\begin{equation}
\frac{\partial\boldsymbol{Q}}{\partial t}+\boldsymbol{M}\left(\boldsymbol{Q}\right)\cdot\nabla\boldsymbol{Q}=\boldsymbol{0}
\end{equation}

where $\boldsymbol{M}=\frac{\partial\boldsymbol{F}}{\partial\boldsymbol{Q}}+\boldsymbol{B}$.
Let $\boldsymbol{n}$ be the normal to the boundary at point $\boldsymbol{s}\in\partial C$.
For the GPR model, $\hat{M}=\boldsymbol{M}\left(\boldsymbol{Q}\left(\boldsymbol{s}\right)\right)\cdot\boldsymbol{n}$
is a diagonalizable matrix with decomposition $\hat{M}=\hat{R}\hat{\Lambda}\hat{R}^{-1}$
where the columns of $\hat{R}$ are the right eigenvectors and $\hat{\Lambda}$
is the diagonal matrix of eigenvalues. Define also $\boldsymbol{\hat{F}}=\boldsymbol{F}\cdot\boldsymbol{n}$
and $\hat{B}=\boldsymbol{B}\cdot\boldsymbol{n}$. Using these definitions,
the interface terms arising in the FV formula have the following form:

\begin{equation}
\boldsymbol{\mathcal{D}}\left(\boldsymbol{Q^{-}},\boldsymbol{Q^{+}}\right)=\frac{1}{2}\left(\boldsymbol{\hat{F}}\left(\boldsymbol{Q^{+}}\right)+\boldsymbol{\hat{F}}\left(\boldsymbol{Q^{-}}\right)+\tilde{B}\left(\boldsymbol{Q^{+}}-\boldsymbol{Q^{-}}\right)+\tilde{M}\left(\boldsymbol{Q^{+}}-\boldsymbol{Q^{-}}\right)\right)
\end{equation}

$\tilde{M}$ is chosen to either correspond to a Rusanov/Lax-Friedrichs
flux (see \citet{Toro2009}):

\begin{equation}
\tilde{M}=\max\left(\max\left|\hat{\Lambda}\left(\boldsymbol{Q^{+}}\right)\right|,\max\left|\hat{\Lambda}\left(\boldsymbol{Q^{-}}\right)\right|\right)
\end{equation}

or a simplified Osher\textendash Solomon flux (see \citet{Dumbser2011a,Dumbser2011b}):

\begin{equation}
\tilde{M}=\int_{0}^{1}\left|\hat{M}\left(\boldsymbol{Q^{-}}+z\left(\boldsymbol{Q^{+}}-\boldsymbol{Q^{-}}\right)\right)\right|dz
\end{equation}

where

\begin{equation}
\left|\hat{M}\right|=\hat{R}\left|\hat{\Lambda}\right|\hat{R}^{-1}
\end{equation}

$\tilde{B}$ takes the following form:

\begin{equation}
\tilde{B}=\int_{0}^{1}\hat{B}\left(\boldsymbol{Q^{-}}+z\left(\boldsymbol{Q^{+}}-\boldsymbol{Q^{-}}\right)\right)dz
\end{equation}

It was found that the Osher-Solomon flux would often produce slightly
less diffusive results, but that it was more computationally expensive,
and also had a greater tendency to introduce numerical artefacts.

$\boldsymbol{P_{i}^{n+\frac{1}{2}}},\boldsymbol{D_{i}^{n+\frac{1}{2}}}$
are calculated using an $N+1$-point Gauss-Legendre quadrature, replacing
$\boldsymbol{Q}\left(\boldsymbol{x},t_{n+\frac{1}{2}}\right)$ with
$\boldsymbol{w^{n+\frac{1}{2}}}\left(\boldsymbol{x}\right)$.

\subsection{The Temporal ODEs}

Noting that $\frac{d\rho}{dt}=0$ over the ODE time step, the operator
$S$ entails solving the following systems:

\begin{subequations}

\begin{align}
\frac{dA}{dt} & =\frac{-3}{\tau_{1}}\left|A\right|^{\frac{5}{3}}A\dev\left(G\right)\label{eq:DistortionODE}\\
\frac{d\boldsymbol{J}}{dt} & =-\frac{1}{\tau_{2}}\frac{T\rho_{0}}{T_{0}\rho}\boldsymbol{J}\label{eq:ThermalODE}
\end{align}

\end{subequations}

These systems can be solved concurrently with a stiff ODE solver.
The Jacobians of these two systems to be used in an ODE solver are
given in \prettyref{subsec:Jacobian-of-Distortion-ODEs} and \prettyref{subsec:Jacobian-of-Thermal-Impulse-ODEs}.
However, these systems can also be solved separately, using the analytical
results presented in \prettyref{sec:GPR-Specific-Performance-Improvements},
under specific assumptions. The second-order Strang splitting is then:

\begin{equation}
\boldsymbol{Q_{\Delta t}}=D^{\frac{\Delta t}{2}}T^{\frac{\Delta t}{2}}H^{\Delta t}T^{\frac{\Delta t}{2}}D^{\frac{\Delta t}{2}}\boldsymbol{Q_{0}}
\end{equation}

where $D^{\delta t},T^{\delta t}$ are the operators solving the distortion
and thermal impulse ODEs respectively, over timestep $\delta t$.
This allows us to bypass the relatively computationally costly process
of solving these systems numerically.

\section{GPR-Specific Performance Improvements\label{sec:GPR-Specific-Performance-Improvements}}

\subsection{The Thermal Impulse ODEs\label{subsec:The-Thermal-Impulse-ODEs}}

Taking the EOS for the GPR model \eqref{eq:EOS} and denoting by $E_{2}^{\left(A\right)},E_{2}^{\left(J\right)}$
the components of $E_{2}$ depending on $A$ and $\boldsymbol{J}$
respectively, we have:

\begin{align}
T & =\frac{E_{1}}{c_{v}}\\
 & =\frac{E-E_{2}^{\left(A\right)}\left(A\right)-E_{3}\left(\boldsymbol{v}\right)}{c_{v}}-\frac{1}{c_{v}}E_{2}^{\left(J\right)}\left(\boldsymbol{J}\right)\nonumber \\
 & =c_{1}-c_{2}\left\Vert \boldsymbol{J}\right\Vert ^{2}\nonumber 
\end{align}

where:

\begin{subequations}

\begin{align}
c_{1} & =\frac{E-E_{2}^{\left(A\right)}\left(A\right)-E_{3}\left(\boldsymbol{v}\right)}{c_{v}}\\
c_{2} & =\frac{\alpha^{2}}{2c_{v}}
\end{align}

\end{subequations}

Over the time period of the ODE \eqref{eq:ThermalODE}, $c_{1},c_{2}>0$
are constant. We have:

\begin{equation}
\frac{dJ_{i}}{dt}=-\left(\frac{1}{\tau_{2}}\frac{\rho_{0}}{T_{0}\rho}\right)J_{i}\left(c_{1}-c_{2}\left\Vert \boldsymbol{J}\right\Vert ^{2}\right)
\end{equation}

Therefore:

\begin{equation}
\frac{d}{dt}\left(J_{i}^{2}\right)=J_{i}^{2}\left(-a+b\left(J_{1}^{2}+J_{2}^{2}+J_{3}^{2}\right)\right)
\end{equation}

where

\begin{subequations}

\begin{align}
a & =\frac{2\rho_{0}}{\tau_{2}T_{0}\rho c_{v}}\left(E-E_{2}^{\left(A\right)}\left(A\right)-E_{3}\left(\boldsymbol{v}\right)\right)\\
b & =\frac{\rho_{0}\alpha^{2}}{\tau_{2}T_{0}\rho c_{v}}
\end{align}

\end{subequations}

Note that this is a generalized Lotka-Volterra system in $\left\{ J_{1}^{2},J_{2}^{2},J_{3}^{2}\right\} $.
It has the following analytical solution:

\begin{equation}
\boldsymbol{J}\left(t\right)=\boldsymbol{J}\left(0\right)\sqrt{\frac{1}{e^{at}-\frac{b}{a}\left(e^{at}-1\right)\left\Vert \boldsymbol{J}\left(0\right)\right\Vert ^{2}}}
\end{equation}

\subsection{The Distortion ODEs}

\subsubsection{Reduced Distortion ODEs}

Let $k_{0}=\frac{3}{\tau_{1}}\left(\frac{\rho}{\rho_{0}}\right)^{\frac{5}{3}}>0$
and let $A$ have singular value decomposition $U\Sigma V^{T}$. Then:

\begin{equation}
G=\left(U\Sigma V^{T}\right)^{T}U\Sigma V^{T}=V\Sigma^{2}V^{T}
\end{equation}

\begin{equation}
\tr\left(G\right)=\tr\left(V\Sigma^{2}V^{T}\right)=\tr\left(\Sigma^{2}V^{T}V\right)=\tr\left(\Sigma^{2}\right)
\end{equation}

Therefore:

\begin{align}
\frac{dA}{dt} & =-k_{0}U\Sigma V^{T}\left(V\Sigma^{2}V^{T}-\frac{\tr\left(\Sigma^{2}\right)}{3}I\right)\\
 & =-k_{0}U\Sigma\left(\Sigma^{2}-\frac{\tr\left(\Sigma^{2}\right)}{3}\right)V^{T}\nonumber \\
 & =-k_{0}U\Sigma\dev\left(\Sigma^{2}\right)V^{T}\nonumber 
\end{align}

It is a common result (see \citet{Giles2008}) that:

\begin{equation}
d\Sigma=U^{T}dAV
\end{equation}

and thus:

\begin{equation}
\frac{d\Sigma}{dt}=-k_{0}\Sigma\dev\left(\Sigma^{2}\right)
\end{equation}

Using a fast $3\times3$ SVD algorithm (such as in \citet{McAdams2011}),
$U,V,\Sigma$ can be obtained, after which the following procedure
is applied to $\Sigma$, giving $A\left(t\right)=U\Sigma\left(t\right)V^{T}$.

Denote the singular values of $A$ by $a_{1},a_{2},a_{3}$. Then:

\begin{equation}
\Sigma\dev\left(\Sigma^{2}\right)=\left(\begin{array}{ccc}
a_{1}\left(a_{1}^{2}-\frac{a_{1}^{2}+a_{2}^{2}+a_{3}^{2}}{3}\right) & 0 & 0\\
0 & a_{1}\left(a_{1}^{2}-\frac{a_{1}^{2}+a_{2}^{2}+a_{3}^{2}}{3}\right) & 0\\
0 & 0 & a_{1}\left(a_{1}^{2}-\frac{a_{1}^{2}+a_{2}^{2}+a_{3}^{2}}{3}\right)
\end{array}\right)
\end{equation}

Letting $x_{i}=\frac{a_{i}^{2}}{\det\left(A\right)^{\frac{2}{3}}}=\frac{a_{i}^{2}}{\left(\frac{\rho}{\rho_{0}}\right)^{\frac{2}{3}}}$
we have:

\begin{equation}
\frac{dx_{i}}{d\tau}=-3x_{i}\left(x_{i}-\bar{x}\right)\label{eq:StretchODESystem}
\end{equation}

where $\tau=\frac{2}{\tau_{1}}\left(\frac{\rho}{\rho_{0}}\right)^{\frac{7}{3}}t$
and $\bar{x}$ is the arithmetic mean of $x_{1},x_{2},x_{3}$. This
ODE system travels along the surface $\Psi=\left\{ x_{1},x_{2},x_{3}>0,x_{1}x_{2}x_{3}=1\right\} $
to the point $x_{1},x_{2},x_{3}=1$. This surface is symmetrical in
the planes $x_{1}=x_{2}$, $x_{1}=x_{3}$, $x_{2}=x_{3}$. As such,
given that the system is autonomous, the paths of evolution of the
$x_{i}$ cannot cross the intersections of these planes with $\Psi$.
Thus, any non-strict inequality of the form $x_{i}\geq x_{j}\geq x_{k}$
is maintained for the whole history of the system. By considering
\eqref{eq:StretchODESystem} it is clear that in this case $x_{i}$
is monotone decreasing, $x_{k}$ is monotone increasing, and the time
derivative of $x_{j}$ may switch sign.

Note that we have:

\begin{equation}
\left\{ \begin{array}{c}
\frac{dx_{i}}{d\tau}=-x_{i}\left(2x_{i}-x_{j}-x_{k}\right)=-x_{i}\left(2x_{i}-x_{j}-\frac{1}{x_{i}x_{j}}\right)\\
\frac{dx_{j}}{d\tau}=-x_{j}\left(2x_{j}-x_{k}-x_{i}\right)=-x_{j}\left(2x_{j}-x_{i}-\frac{1}{x_{i}x_{j}}\right)
\end{array}\right.\label{eq:2odes}
\end{equation}

Thus, an ODE solver can be used on these two equations to effectively
solve the ODEs for all 9 components of $A$. Note that:

\begin{equation}
\frac{dx_{j}}{dx_{i}}=\frac{x_{j}}{x_{i}}\frac{2x_{j}-x_{i}-\frac{1}{x_{i}x_{j}}}{2x_{i}-x_{j}-\frac{1}{x_{i}x_{j}}}
\end{equation}

This has solution:

\begin{equation}
x_{j}=\frac{c+\sqrt{c^{2}+4\left(1-c\right)x_{i}^{3}}}{2x_{i}^{2}}
\end{equation}

where

\begin{equation}
c=-\frac{x_{i,0}\left(x_{i,0}x_{j,0}^{2}-1\right)}{x_{i,0}-x_{j,0}}\in\left(-\infty,0\right]
\end{equation}

In the case that $x_{i,0}=x_{j,0}$, we have $x_{i}=x_{j}$ for all
time. Thus, the ODE system for $A$ has been reduced to a single ODE,
as $x_{j}\left(x_{i}\right)$ can be inserted into the RHS of the
equation for $\frac{dx_{i}}{d\tau}$. However, it is less computationally
expensive to evolve the system presented in \eqref{eq:2odes}.

\subsubsection{Bounds on Reduced Distortion ODEs}

If any of the relations in $x_{i}\geq x_{j}\geq x_{k}$ are in fact
equalities, equality is maintained throughout the history of the system.
This can be seen by noting that the time derivatives of the equal
variables are in this case equal. If $x_{j}=x_{k}$ then $x_{i}=\frac{1}{x_{j}^{2}}$.
Combining these results, the path of the system in $\left(x_{i},x_{j}\right)$
coordinates is in fact confined to the curved triangular region:

\begin{equation}
\left\{ \left(x_{i},x_{j}\right):x_{i}\leq x_{i}^{0}\:\cap\:x_{i}\geq x_{j}\:\cap\:x_{i}\geq\frac{1}{x_{j}^{2}}\right\} 
\end{equation}

\begin{figure}
\begin{centering}
\includegraphics[width=0.5\textwidth]{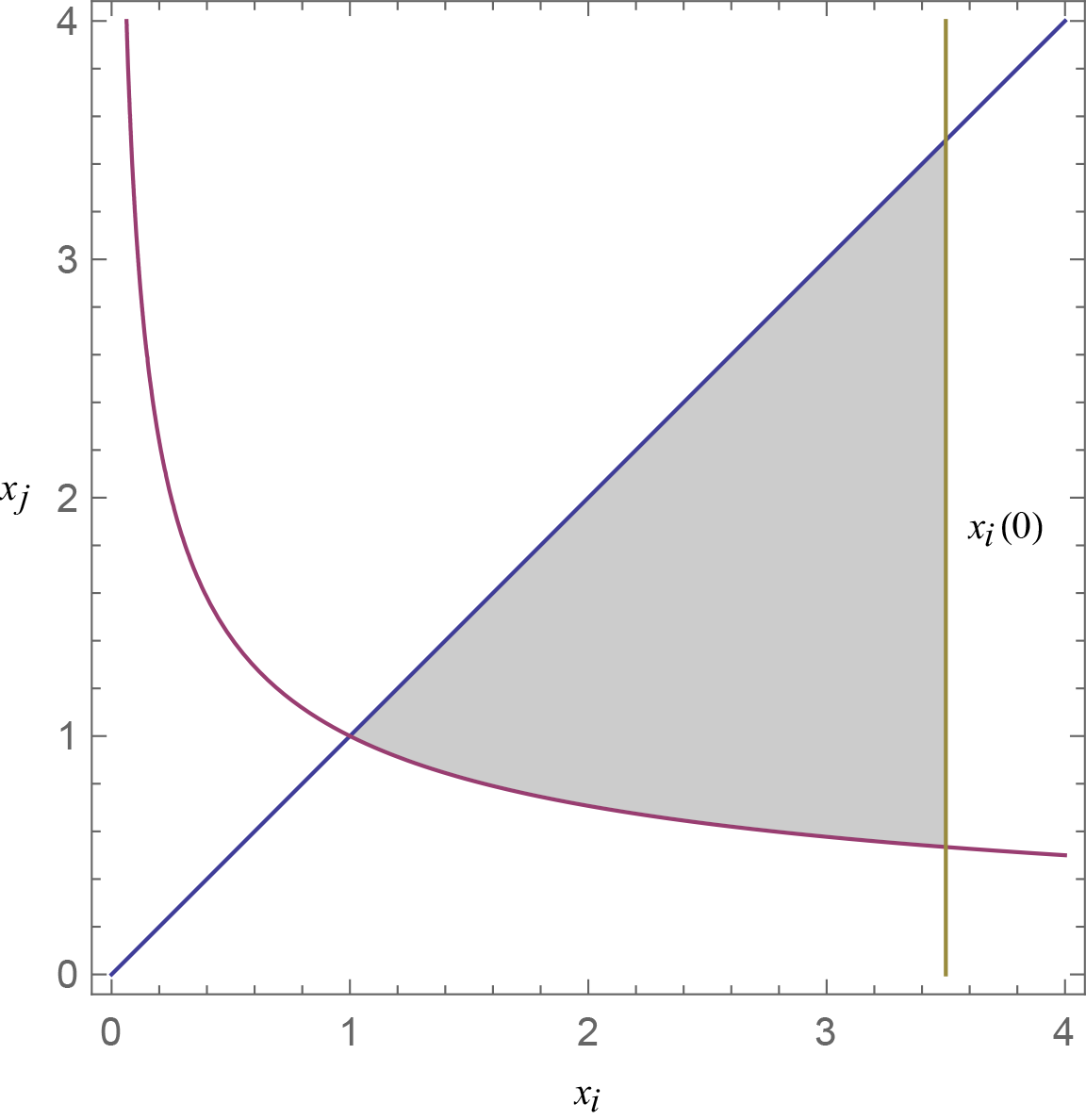}
\par\end{centering}
\caption{\label{fig:EquationBounds}The (shaded) region to which $x_{i},x_{j}$
are confined in the evolution of the distortion ODEs}
\end{figure}

This is demonstrated in \prettyref{fig:EquationBounds}. By \eqref{eq:2odes},
the rate of change of $x_{i}$ at a particular value $x_{i}=x_{i}^{*}$
is given by:

\begin{equation}
-x_{i}^{*}\left(2x_{i}^{*}-x_{j}-\frac{1}{x_{i}^{*}x_{j}}\right)
\end{equation}

Note that:

\begin{align}
\frac{d}{dx_{j}}\left(2x_{i}^{*}-x_{j}-\frac{1}{x_{i}^{*}x_{j}}\right) & =-1+\frac{1}{x_{i}^{*}x_{j}^{2}}=0\\
 & \Rightarrow x_{j}=\frac{1}{\sqrt{x_{i}^{*}}}\nonumber 
\end{align}

\begin{equation}
\frac{d^{2}}{dx_{j}^{2}}\left(2x_{i}^{*}-x_{j}-\frac{1}{x_{i}^{*}x_{j}}\right)=\frac{-2}{x_{i}^{*}x_{j}^{3}}<0
\end{equation}

Thus, $x_{i}$ decreases fastest on the line $x_{i}=\frac{1}{x_{j}^{2}}$
(the bottom boundary of the region given in \prettyref{fig:EquationBounds}),
and slowest on the line $x_{i}=x_{j}$. The rates of change of $x_{i}$
along these two lines are given respectively by:

\begin{subequations}

\begin{align}
\frac{dx_{i}}{d\tau} & =-2x_{i}\left(x_{i}-\sqrt{\frac{1}{x_{i}}}\right)\\
\frac{dx_{i}}{d\tau} & =-x_{i}\left(x_{i}-\frac{1}{x_{i}^{2}}\right)
\end{align}

\end{subequations}

These have implicit solutions:

\begin{subequations}

\begin{align}
\tau & =\left(f\left(\sqrt{x_{i}}\right)+g\left(\sqrt{x_{i}}\right)\right)-\left(f\left(\sqrt{x_{i}^{0}}\right)+g\left(\sqrt{x_{i}^{0}}\right)\right)\equiv F_{1}\left(x_{i};x_{i}^{0}\right)\\
\tau & =\left(f\left(x_{i}\right)-g\left(x_{i}\right)\right)-\left(f\left(x_{i}^{0}\right)-g\left(x_{i}^{0}\right)\right)\equiv F_{2}\left(x_{i};x_{i}^{0}\right)
\end{align}

\end{subequations}

where

\begin{subequations}

\begin{align}
f\left(x_{i}\right) & =\frac{1}{6}\log\left(\frac{x_{i}^{2}+x_{i}+1}{\left(x_{i}-1\right)^{2}}\right)\\
g\left(x_{i}\right) & =\frac{1}{\sqrt{3}}\tan^{-1}\left(\frac{2x_{i}+1}{\sqrt{3}}\right)
\end{align}

\end{subequations}

As \eqref{eq:StretchODESystem} is an autonomous system of ODEs, it
has the property that its limit $x_{1}=x_{2}=x_{3}=1$ is never obtained
in finite time, in precise arithmetic. In floating point arithmetic
we may say that the system has converged when $x_{i}-1<\epsilon$
(machine epsilon) for each $i$. This happens when:

\begin{equation}
\tau>F_{2}\left(1+\epsilon;x_{i}^{0}\right)
\end{equation}
This provides a quick method to check whether it is necessary to run
the ODE solver in a particular cell. If the following condition is
satisfied then we know the system in that cell converges to the ground
state over the time interval in which the ODE system is calculated:

\begin{equation}
\frac{2}{\tau_{1}}\left(\frac{\rho}{\rho_{0}}\right)^{\frac{7}{3}}\Delta t>F_{2}\left(1+\epsilon;\max\left\{ x_{i}^{0}\right\} \right)
\end{equation}

If the fluid is very inviscid, resulting in a stiff ODE, the critical
time is lower, and there is more chance that the ODE system in the
cell reaches its limit in $\Delta t$. This check potentially saves
a lot of computationally expensive stiff ODE solves. The same goes
for if the flow is slow-moving, as the system will be closer to its
ground state at the start of the time step and is more likely to converge
over $\Delta t$. Similarly, if the following condition is satisfied
then we know for sure that an ODE solver is necessary, as the system
certainly will not have converged over the timestep:

\begin{equation}
\frac{2}{\tau_{1}}\left(\frac{\rho}{\rho_{0}}\right)^{\frac{7}{3}}\Delta t<F_{1}\left(1+\epsilon;\max\left\{ x_{i}^{0}\right\} \right)
\end{equation}

\subsubsection{Analytical Approximation\label{subsec:Analytical-Approximation}}

We now explore cases when even the reduced ODE system \eqref{eq:2odes}
need not be solved numerically. Define the following variables:

\begin{subequations}

\begin{align}
m & =\frac{x_{1}+x_{2}+x_{3}}{3}\\
u & =\frac{\left(x_{1}-x_{2}\right)^{2}+\left(x_{2}-x_{3}\right)^{2}+\left(x_{3}-x_{1}\right)^{2}}{3}
\end{align}

\end{subequations}

It is a standard result that $m\geq\sqrt[3]{x_{1}x_{2}x_{3}}$. Thus,
$m\geq1$. Note that $u$ is proportional to the internal energy contribution
from the distortion. From \eqref{eq:StretchODESystem} we have:

\begin{subequations}

\begin{align}
\frac{du}{d\tau} & =-18\left(1-m\left(m^{2}-\frac{5}{6}u\right)\right)\\
\frac{dm}{d\tau} & =-u
\end{align}

\end{subequations}

Combining these equations, we have:

\begin{equation}
\frac{d^{2}m}{d\tau^{2}}=-\frac{du}{d\tau}=18\left(1-m\left(m^{2}-\frac{5}{6}u\right)\right)
\end{equation}

Therefore:

\begin{equation}
\left\{ \begin{array}{c}
\frac{d^{2}m}{d\tau^{2}}+15m\frac{dm}{d\tau}+18\left(m^{3}-1\right)=0\\
m\left(0\right)=m_{0}\\
m^{'}\left(0\right)=-u_{0}
\end{array}\right.
\end{equation}

We make the following assumption, noting that it is true in all physical
situations tested in this study:

\begin{equation}
m\left(t\right)=1+\eta\left(t\right),\quad\eta\ll1\;\forall t\geq0\label{eq:Assumption}
\end{equation}

Thus, we have the linearized ODE:

\begin{equation}
\left\{ \begin{array}{c}
\frac{d^{2}\eta}{d\tau^{2}}+15\frac{d\eta}{d\tau}+54\eta=0\\
\eta\left(0\right)=m_{0}-1\\
\eta^{'}\left(0\right)=-u_{0}
\end{array}\right.
\end{equation}

This is a Sturm-Liouville equation with solution:

\begin{equation}
\eta\left(\tau\right)=\frac{e^{-9\tau}}{3}\left(\left(9m_{0}-u_{0}-9\right)e^{3\tau}-\left(6m_{0}-u_{0}-6\right)\right)
\end{equation}

Thus, we also have:

\begin{equation}
u\left(\tau\right)=e^{-9\tau}\left(e^{3\tau}\left(18m_{0}-2u_{0}-18\right)-\left(18m_{0}-3u_{0}-18\right)\right)
\end{equation}

Once $m_{\Delta t}=1+\eta\left(\frac{2}{\tau_{1}}\left(\frac{\rho}{\rho_{0}}\right)^{\frac{7}{3}}\Delta t\right)$
and $u_{\Delta t}=u\left(\frac{2}{\tau_{1}}\left(\frac{\rho}{\rho_{0}}\right)^{\frac{7}{3}}\Delta t\right)$
have been found, we have:

\begin{subequations}

\begin{align}
\frac{x_{i}+x_{j}+x_{k}}{3} & =m_{\Delta t}\\
\frac{\left(x_{i}-x_{j}\right)^{2}+\left(x_{j}-x_{k}\right)^{2}+\left(x_{k}-x_{i}\right)^{2}}{3} & =u_{\Delta t}\\
x_{i}x_{j}x_{k} & =1
\end{align}

\end{subequations}

This gives:

\begin{subequations}

\begin{align}
x_{i} & =\frac{\sqrt[3]{6\left(\sqrt{81\Delta^{2}-6u_{\Delta t}^{3}}+9\Delta\right)}}{6}+\frac{u_{\Delta t}}{\sqrt[3]{6\left(\sqrt{81\Delta^{2}-6u_{\Delta t}^{3}}+9\Delta\right)}}+m_{\Delta t}\\
x_{j} & =\frac{1}{2}\left(\sqrt{\frac{x_{i}\left(3m_{\Delta t}-x_{i}\right)^{2}-4}{x_{i}}}+3m_{\Delta t}-x_{i}\right)\\
x_{k} & =\frac{1}{x_{i}x_{j}}
\end{align}

\end{subequations}

where

\begin{equation}
\Delta=-2m_{\Delta t}^{3}+m_{\Delta t}u_{\Delta t}+2
\end{equation}

Note that taking the real parts of the above expression for $x_{i}$
gives:

\begin{subequations}

\begin{align}
x_{i} & =\frac{\sqrt{6u_{\Delta t}}}{3}\cos\left(\frac{\theta}{3}\right)+m_{\Delta t}\\
\theta & =\tan^{-1}\left(\frac{\sqrt{6u_{\Delta t}^{3}-81\Delta^{2}}}{9\Delta}\right)
\end{align}

\end{subequations}

At this point it is not clear which values of $\left\{ x_{i},x_{j},x_{k}\right\} $
are taken by $x_{1},x_{2},x_{3}$. However, this can be inferred from
the fact that any relation $x_{i}\geq x_{j}\geq x_{k}$ is maintained
over the lifetime of the system. Thus, the stiff ODE solver has been
obviated by a few arithmetic operations.

\section{Numerical Results}

\subsection{Strain Relaxation}

In this section, the approximate analytic solver for the distortion
ODEs, presented in \prettyref{subsec:Analytical-Approximation}, is
compared with a numerical ODE solver. Initial data was taken from
\citet{Barton2010}:

\begin{equation}
A=\left(\begin{array}{ccc}
1 & 0 & 0\\
-0.01 & 0.95 & 0.02\\
-0.015 & 0 & 0.9
\end{array}\right)^{-1}
\end{equation}

\begin{figure}[p]
\begin{centering}
\includegraphics[width=0.5\textwidth]{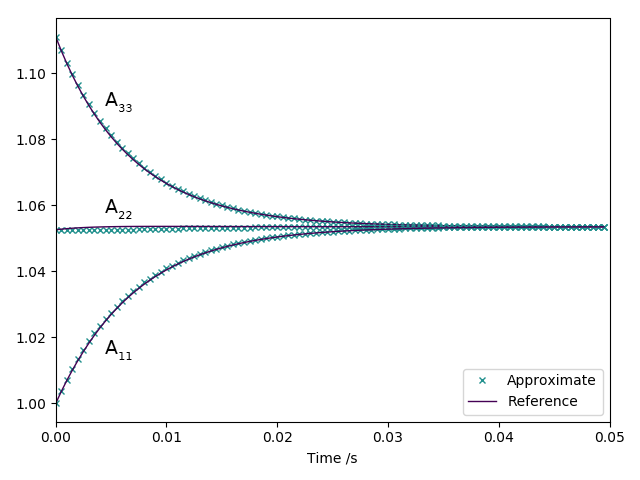}\includegraphics[width=0.5\textwidth]{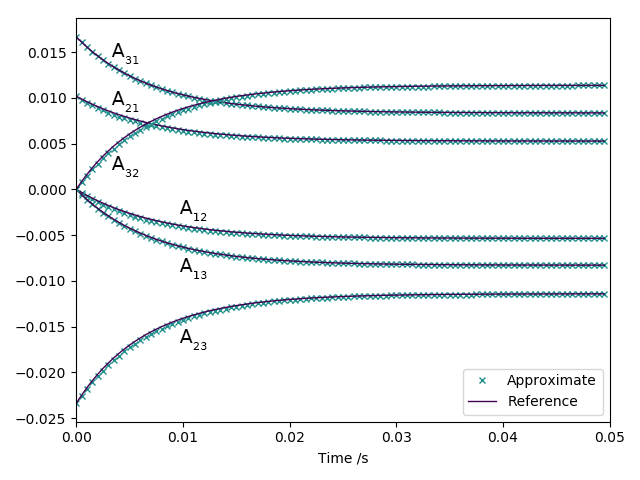}
\par\end{centering}
\caption{\label{fig:Distortion-ODEs_A}The components of the distortion tensor
in the Strain Relaxation Test}
\end{figure}

\begin{figure}[p]
\begin{centering}
\includegraphics[width=0.5\textwidth]{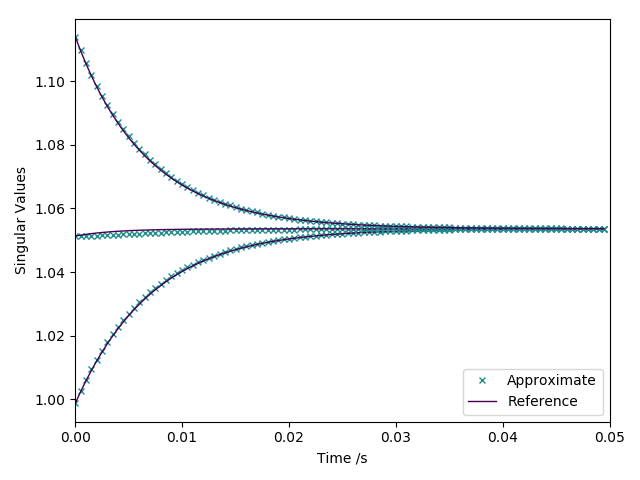}\includegraphics[width=0.5\textwidth]{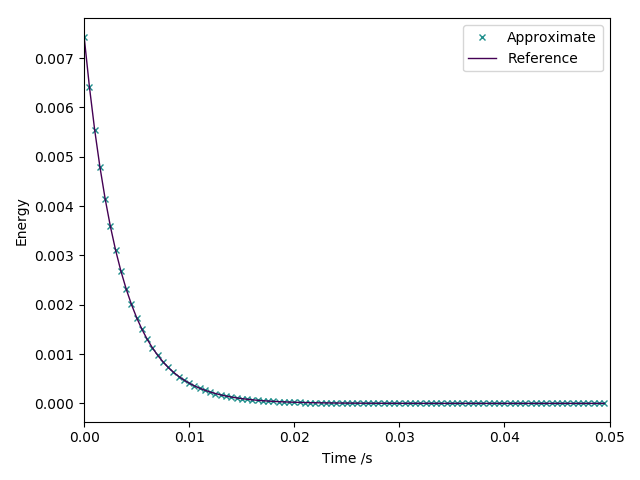}
\par\end{centering}
\caption{\label{fig:Distortion-ODEs_E}The singular values of the distortion
tensor and the energy in the Strain Relaxation Test}
\end{figure}

\begin{figure}[p]
\begin{centering}
\includegraphics[width=0.5\textwidth]{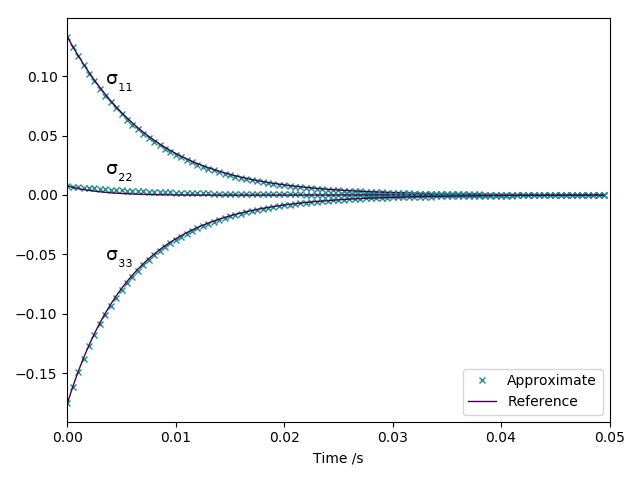}\includegraphics[width=0.5\textwidth]{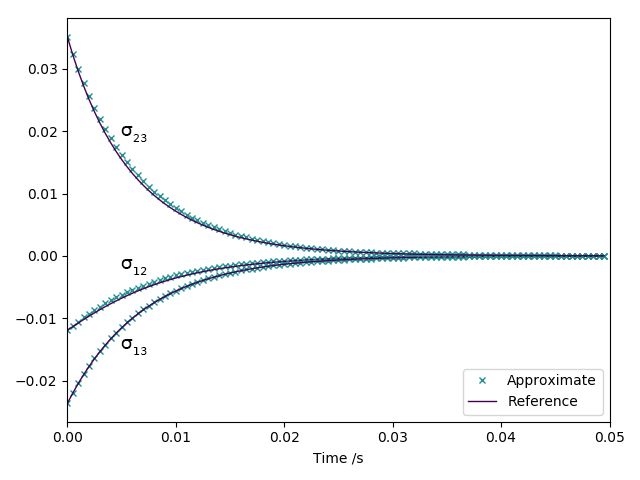}
\par\end{centering}
\caption{\label{fig:Distortion-ODEs_=0003C3}The components of the stress tensor
in the Strain Relaxation Test}
\end{figure}

Additionally, the following parameter values were used: $\rho_{0}=1,c_{s}=1,\mu=10^{-2}$,
giving $\tau_{1}=0.06$. As can be seen in \prettyref{fig:Distortion-ODEs_A},
\prettyref{fig:Distortion-ODEs_E}, and \prettyref{fig:Distortion-ODEs_=0003C3},
the approximate analytic solver compares well with the numerical solver
in its results for the distortion tensor $A$, and thus also the internal
energy and stress tensor. The numerical ODE solver was the odeint
solver from SciPy 0.18.1, based on the LSODA solver from the FORTRAN
library ODEPACK (see \citet{Jones}).

\subsection{Stokes' First Problem\label{subsec:Stokes'-First-Problem}}

This problem is one of the few test cases with an analytic solution
for the Navier-Stokes equations. It consists of two ideal gases in
an infinite domain, meeting at the plane $x=0$, initially flowing
with equal and opposite velocity $\pm0.1$ in the $y$-axis. The initial
conditions are given in \prettyref{tab:SlowOpposingShearFlow}.

\begin{table}
\begin{centering}
\bigskip{}
\begin{tabular}{|c|c|c|c|c|c|}
\hline 
 &
$\rho$ &
$p$ &
\textbf{$\boldsymbol{v}$} &
$A$ &
\textbf{$\boldsymbol{J}$}\tabularnewline
\hline 
\hline 
$x<0$ &
$1$ &
$1/\gamma$ &
$\left(0,-0.1,0\right)$ &
$I_{3}$ &
$\boldsymbol{0}$\tabularnewline
\hline 
$x\geq0$ &
$1$ &
$1/\gamma$ &
$\left(0,0.1,0\right)$ &
$I_{3}$ &
$\boldsymbol{0}$\tabularnewline
\hline 
\end{tabular}
\par\end{centering}
\caption{\label{tab:SlowOpposingShearFlow}Initial conditions for the slow
opposing shear flow test}
\medskip{}
\end{table}

The flow has a low Mach number of 0.1, and this test case is designed
to demonstrate the efficacy of the numerical methods in this flow
regime. The exact solution to the Navier-Stokes equations is given
by\footnote{In this problem, the Navier-Stokes equations reduce to $v_{t}=\mu v_{xx}$.
Defining $\eta=\frac{x}{2\sqrt{\mu t}}$, and assuming $v=f\left(\eta\right)$,
this becomes $f^{''}+2\eta f^{'}=0$. The result follows by solving
this equation with the boundary conditions $v\left(\pm\infty\right)=\pm v_{0}$.}:

\begin{equation}
v=v_{0}\erf\left(\frac{x}{2\sqrt{\mu t}}\right)
\end{equation}

Heat conduction is neglected, and $\gamma=1.4$, $c_{v}=1$, $\rho_{0}=1$,
$c_{s}=1$. The viscosity is variously taken to be $\mu=10^{-2}$,
$\mu=10^{-3}$, $\mu=10^{-4}$ (resulting in $\tau_{1}=0.06$, $\tau_{1}=0.006$,
$\tau_{1}=0.0006$, respectively). Due to the stiffness of the source
terms in the equations governing $A$ in the case that $\mu=10^{-4}$,
the step \eqref{eq:WENO half step} in the WENO reconstruction under
the Split-WENO method was not performed, and $\boldsymbol{w_{p}^{n+\frac{1}{2}}}\equiv\boldsymbol{w_{p}^{n}}$
was taken instead. This avoided the numerical diffusion that otherwise
would have emerged at the interface at $x=0$.

\begin{figure}[p]
\begin{centering}
\includegraphics[width=0.5\textwidth]{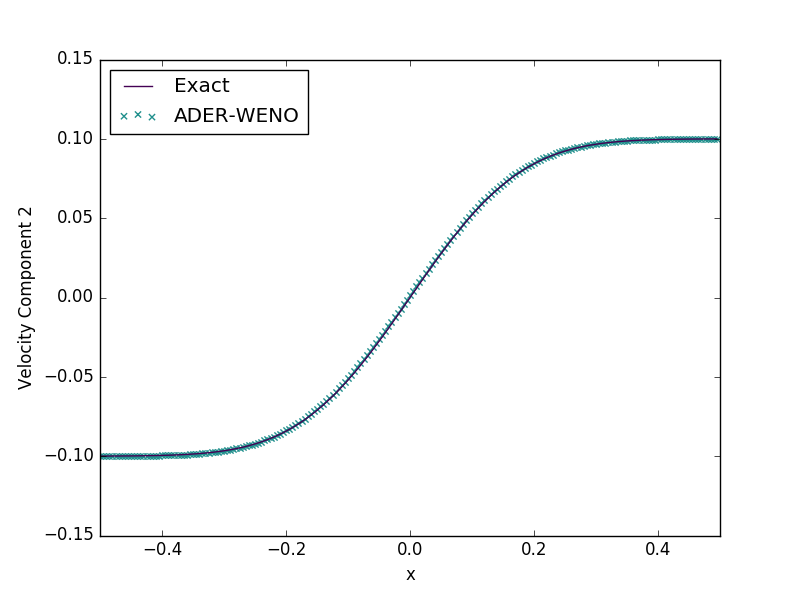}\includegraphics[width=0.5\textwidth]{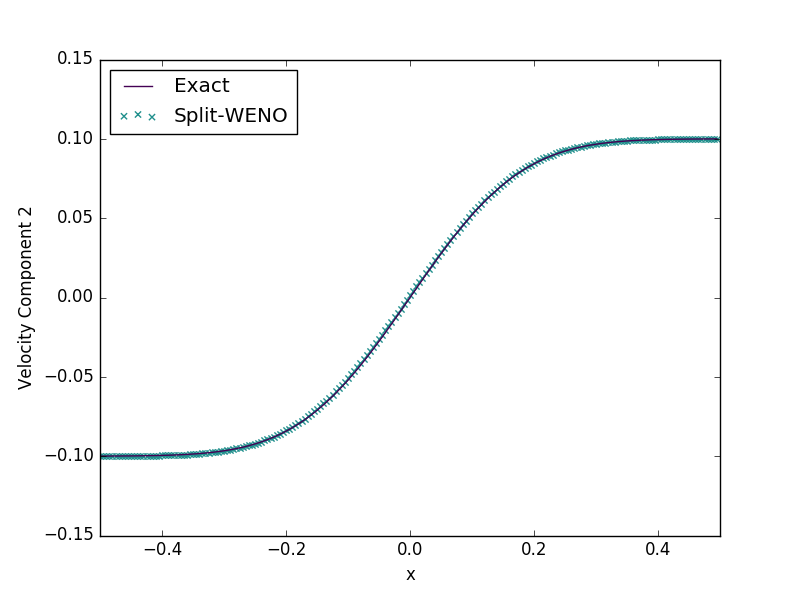}
\par\end{centering}
\begin{centering}
\includegraphics[width=0.5\textwidth]{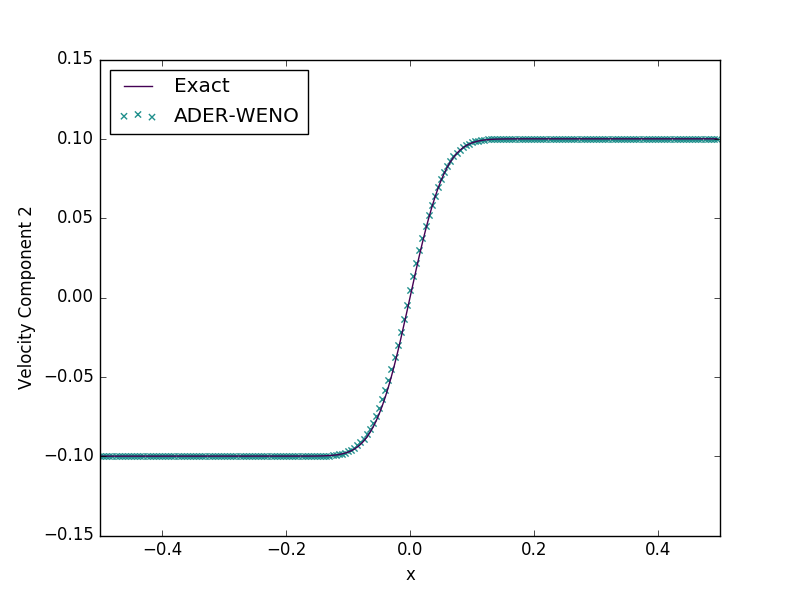}\includegraphics[width=0.5\textwidth]{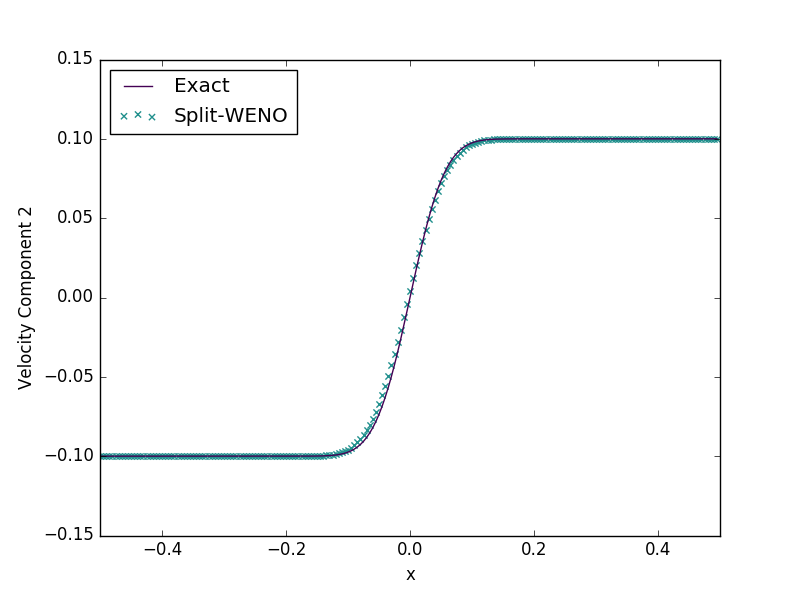}
\par\end{centering}
\begin{centering}
\includegraphics[width=0.5\textwidth]{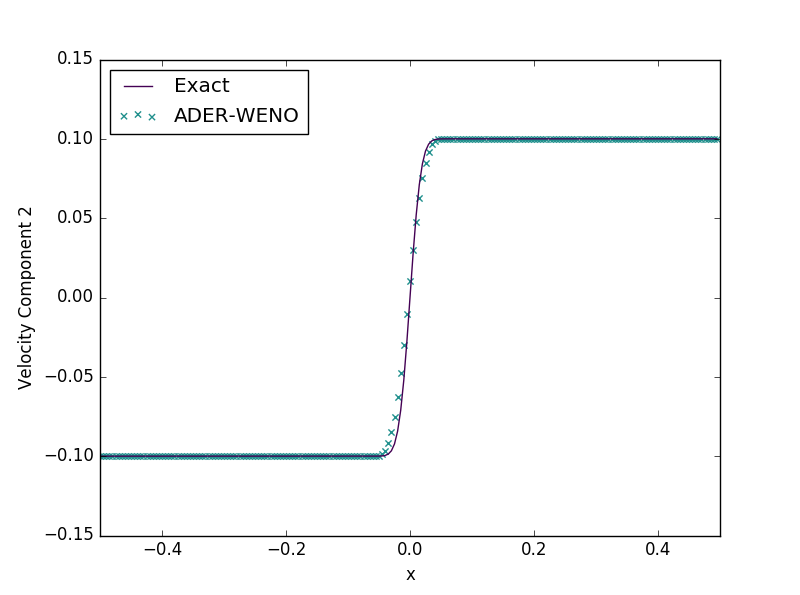}\includegraphics[width=0.5\textwidth]{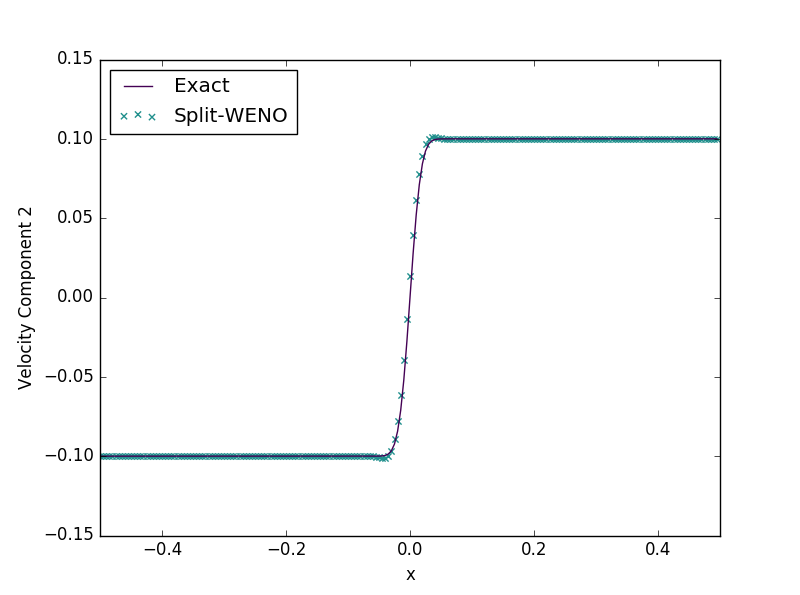}
\par\end{centering}
\caption{\label{fig:Stokes}Results of solving Stokes' First Problem ($\mu=10^{-2},\mu=10^{-3},\mu=10^{-4}$)
with an ADER-WENO scheme and a Split-WENO scheme ($N=2$)}
\end{figure}

The results of simulations with 200 cells at time $t=1$, using reconstruction
polynomials of order $N=2$, are presented in \prettyref{fig:Stokes}.
The GPR model solved with both the ADER-WENO and Split-WENO methods
closely matches the exact Navier-Stokes solution. Note that at $\mu=10^{-2}$
and $\mu=10^{-3}$, the ADER-WENO and Split-WENO methods are almost
indistinguishable. At $\mu=10^{-4}$ the Split-WENO method matches
the curve of the velocity profile more closely, but overshoots slightly
at the boundaries of the center region. This overshoot phenomenon
is not visible in the ADER-WENO results.

\subsection{Viscous Shock}

This test is designed to demonstrate that the numerical methods used
are also able to cope with fast flows. First demonstrated by Becker
\citet{Beker1929}, the Navier-Stokes equations have an analytic solution
for $P_{r}=0.75$ (see Johnson \citet{Johnson2013} for a full analysis).
As noted by Dumbser \citet{Dumbser2016a}, if the wave has nondimensionalised
upstream velocity $\bar{v}=1$ and Mach number $M_{c}$, then its
nondimensionalised downstream velocity is:

\begin{equation}
a=\frac{1+\frac{\gamma-1}{2}M_{c}^{2}}{\frac{\gamma+1}{2}M_{c}^{2}}
\end{equation}

The wave's velocity profile $\bar{v}\left(x\right)$ is given by the
roots of the following equation:

\begin{subequations}

\begin{align}
\frac{1-\bar{v}}{\left(\bar{v}-a\right)^{a}} & =c_{1}\exp\left(-c_{2}x\right)\\
c_{1} & =\left(\frac{1-a}{2}\right)^{1-a}\\
c_{2} & =\frac{3}{4}R_{e}\frac{M_{c}^{2}-1}{\gamma M_{c}^{2}}
\end{align}

\end{subequations}

$c_{1},c_{2}$ are constants that affect the position of the center
of the wave, and its stretch factor, respectively. Following the analysis
of Morduchow and Libby \citet{Morduchow}, the nondimensional pressure
and density profiles are given by:

\begin{equation}
\bar{p}=\frac{1}{\bar{v}}\left(1+\frac{\gamma-1}{2}M_{c}^{2}\left(1-\bar{v}^{2}\right)\right)
\end{equation}

\begin{equation}
\bar{\rho}=\frac{1}{\bar{v}}
\end{equation}

To obtain an unsteady shock traveling into a region at rest, a constant
velocity field $v=M_{c}c_{0}$ is imposed on the traveling wave solution
presented here (where $c_{0}$ is the adiabatic sound speed). Thus,
if $p_{0},\rho_{0}$ are the downstream (reference) values for pressure
and density:

\begin{subequations}

\begin{align}
v & =Mc_{0}\left(1-\bar{v}\right)\\
p & =p_{0}\bar{p}\\
\rho & =\rho_{0}\bar{\rho}
\end{align}

\end{subequations}

These functions are used as initial conditions, along with $A=\sqrt[3]{\bar{\rho}}I$
and $\boldsymbol{J}=\boldsymbol{0}$. The downstream density and pressure
are taken to be $\rho_{0}=1$ and $p_{0}=\frac{1}{\gamma}$ (so that
$c_{0}=1$). $M_{c}=2$ and $R_{e}=100$. The material parameters
are taken to be: $\gamma=1.4$, $p_{\infty}=0$, $c_{v}=2.5$, $c_{s}=5$,
$\alpha=5$, $\mu=2\times10^{-2}$, $\kappa=\frac{28}{3}\times10^{-2}$
(resulting in $\tau_{1}=0.0048$, $\tau_{2}=0.00522\dot{6}$).

\begin{figure}[p]
\begin{centering}
\includegraphics[width=0.5\textwidth]{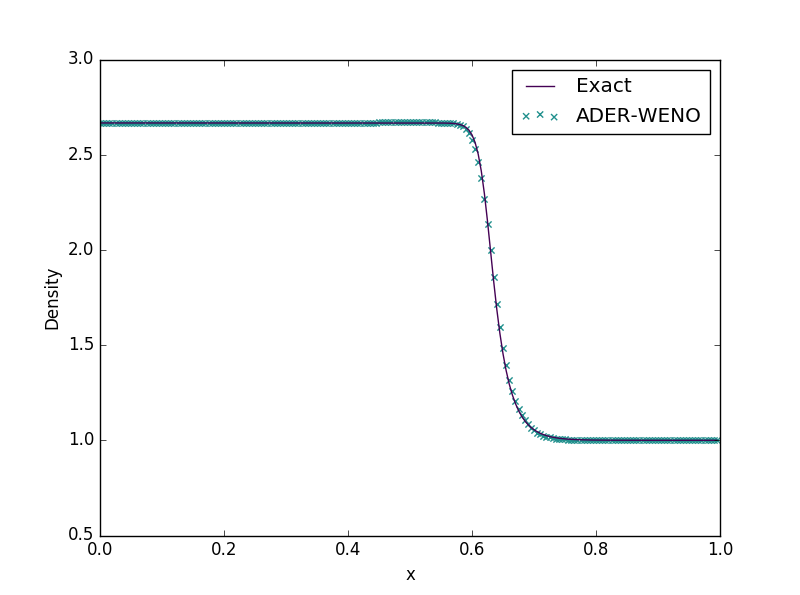}\includegraphics[width=0.5\textwidth]{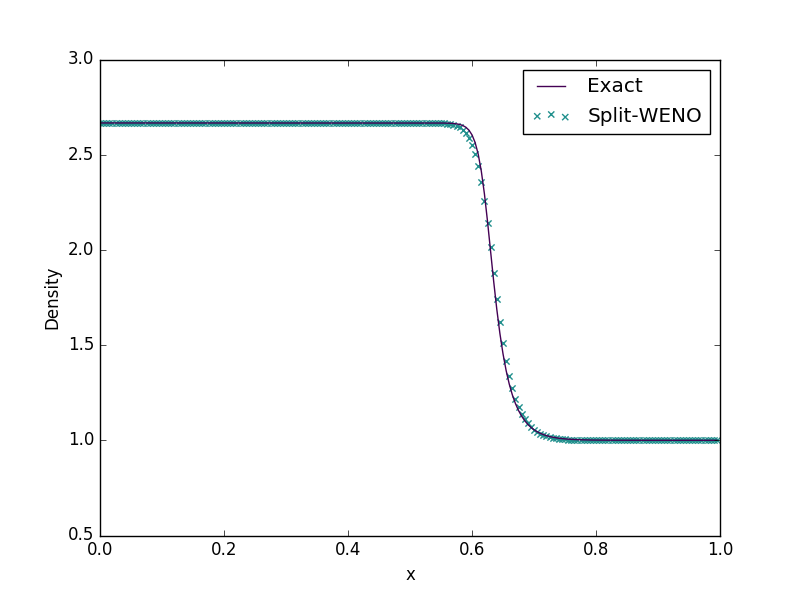}
\par\end{centering}
\begin{centering}
\includegraphics[width=0.5\textwidth]{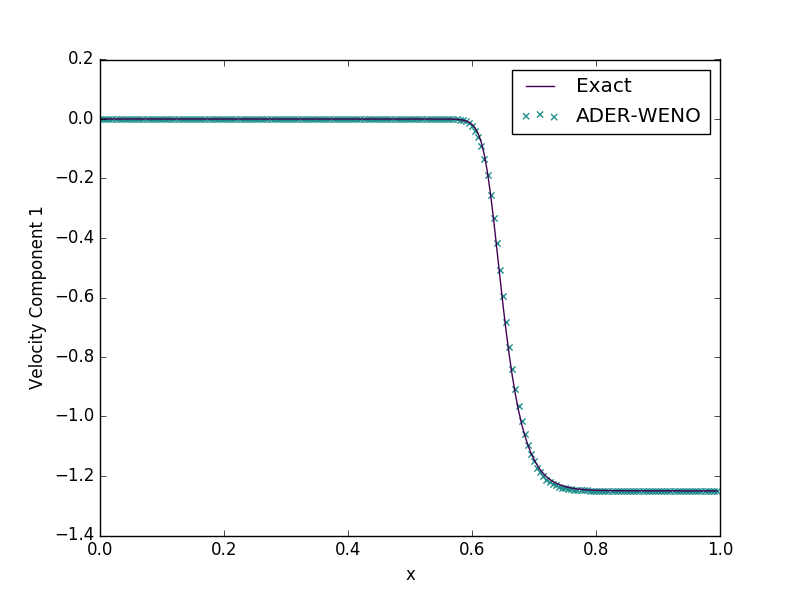}\includegraphics[width=0.5\textwidth]{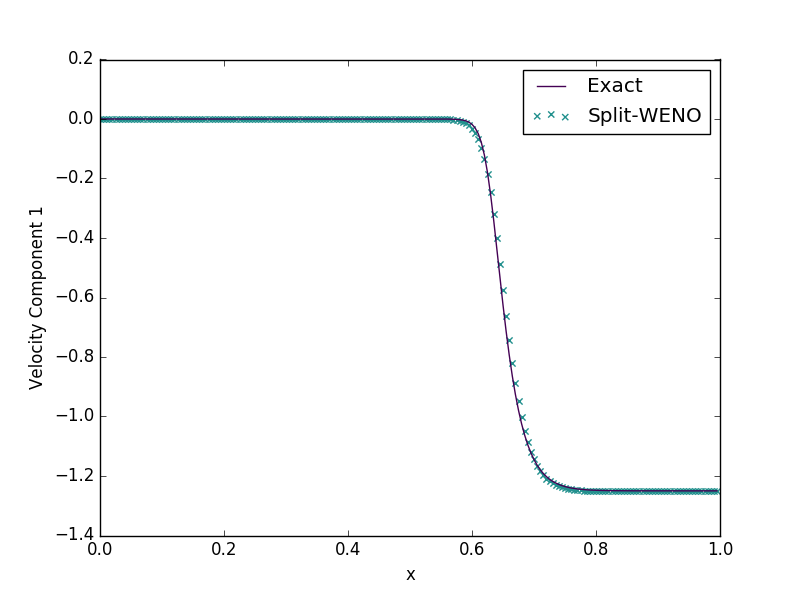}
\par\end{centering}
\begin{centering}
\includegraphics[width=0.5\textwidth]{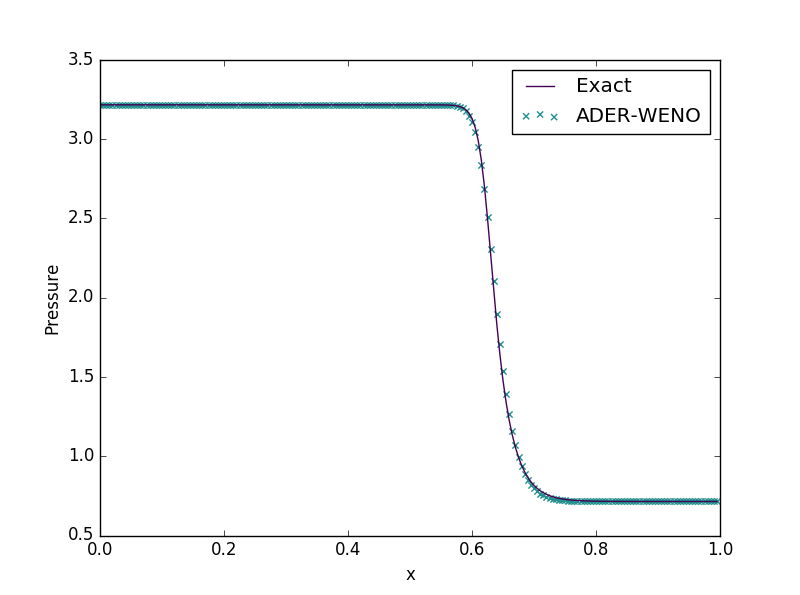}\includegraphics[width=0.5\textwidth]{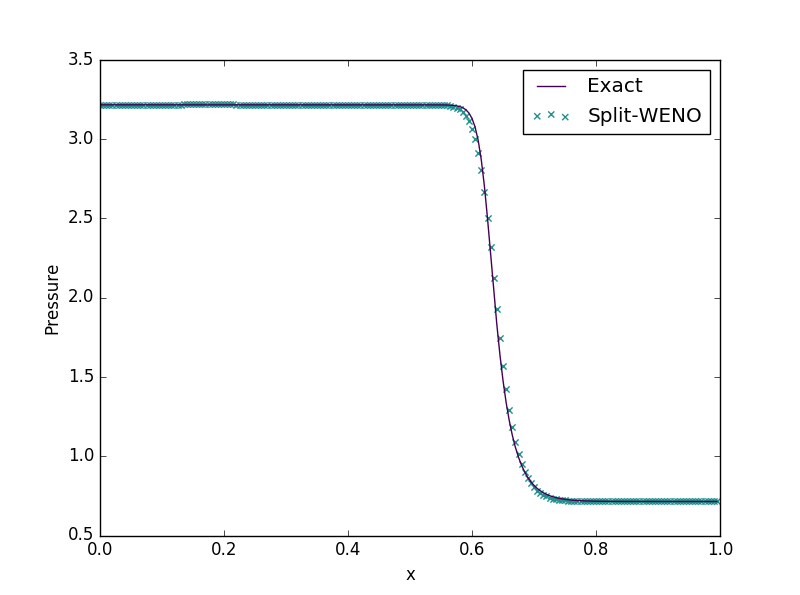}
\par\end{centering}
\caption{\label{fig:=0003C1-v-p Viscous Shock}Density, velocity, and pressure
for the Viscous Shock problem, solved with an ADER-WENO scheme and
a Split-WENO scheme ($N=2$)}
\end{figure}

\begin{figure}[p]
\begin{centering}
\includegraphics[width=0.5\textwidth]{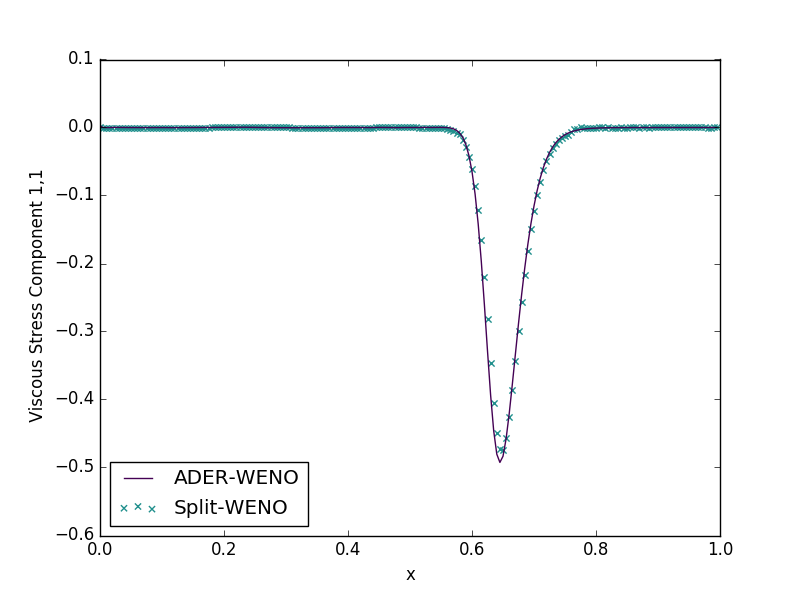}\includegraphics[width=0.5\textwidth]{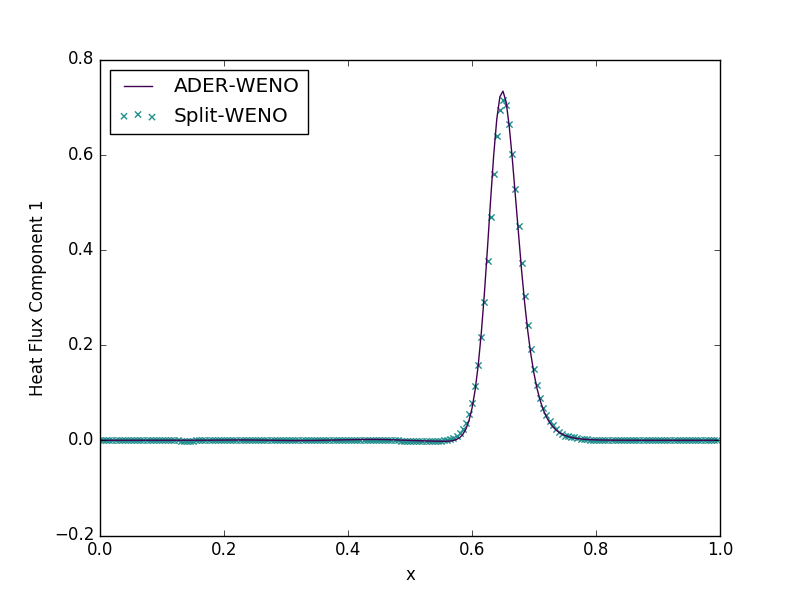}
\par\end{centering}
\caption{\label{fig:=0003C3-q Viscous Shock}Viscous stress and heat flux for
the Viscous Shock problem, solved with both an ADER-WENO scheme and
a Split-WENO scheme ($N=2$)}
\end{figure}

The results of a simulation with 200 cells at time $t=0.2$, using
reconstruction polynomials of order $N=2$, are presented in \prettyref{fig:=0003C1-v-p Viscous Shock}
and \prettyref{fig:=0003C3-q Viscous Shock}. The shock was initially
centered at $x=0.25$, reaching $x=0.65$ at the final time. Note
that the density, velocity, and pressure results for both methods
match the exact solution well, with the ADER-WENO method appearing
to produce a slightly more accurate solution. The results for the
two methods for the stress tensor and heat flux are close.


\subsection{Heat Conduction in a Gas}

This is a simple test case to ensure that the heat transfer terms
in the implementation are working correctly. Two ideal gases at different
temperatures are initially in contact at position $x=0$. The initial
conditions for this problem are given in \prettyref{tab:HeatConduction}.

\begin{table}
\begin{centering}
\bigskip{}
\begin{tabular}{|c|c|c|c|c|c|}
\hline 
 &
$\rho$ &
$p$ &
\textbf{$\boldsymbol{v}$} &
$A$ &
\textbf{$\boldsymbol{J}$}\tabularnewline
\hline 
\hline 
$x<0$ &
$2$ &
$1$ &
$\boldsymbol{0}$ &
$\sqrt[3]{2}\cdot I_{3}$ &
$\boldsymbol{0}$\tabularnewline
\hline 
$x\geq0$ &
$0.5$ &
$1$ &
$\boldsymbol{0}$ &
$\frac{1}{\sqrt[3]{2}}\cdot I_{3}$ &
$\boldsymbol{0}$\tabularnewline
\hline 
\end{tabular}
\par\end{centering}
\caption{\label{tab:HeatConduction}Initial conditions for the heat conduction
test}
\medskip{}
\end{table}

\begin{figure}[p]
\begin{centering}
\includegraphics[width=0.5\textwidth]{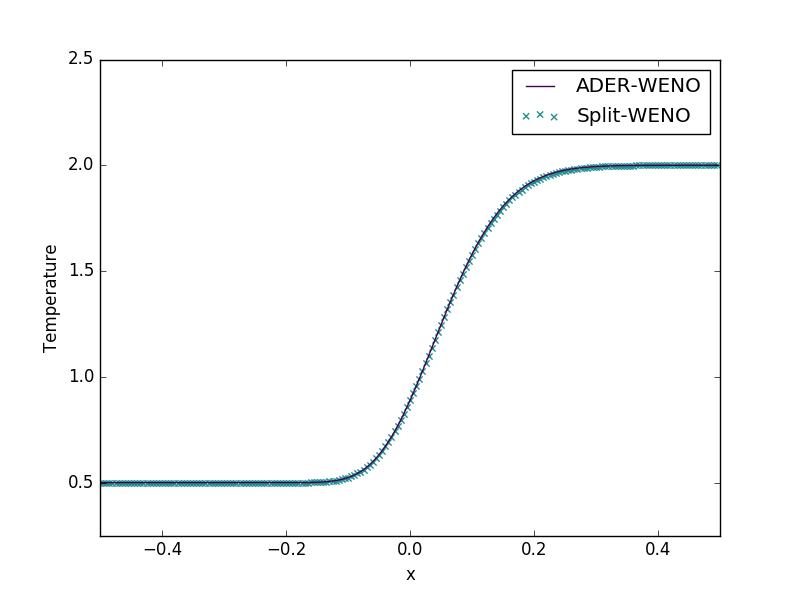}\includegraphics[width=0.5\textwidth]{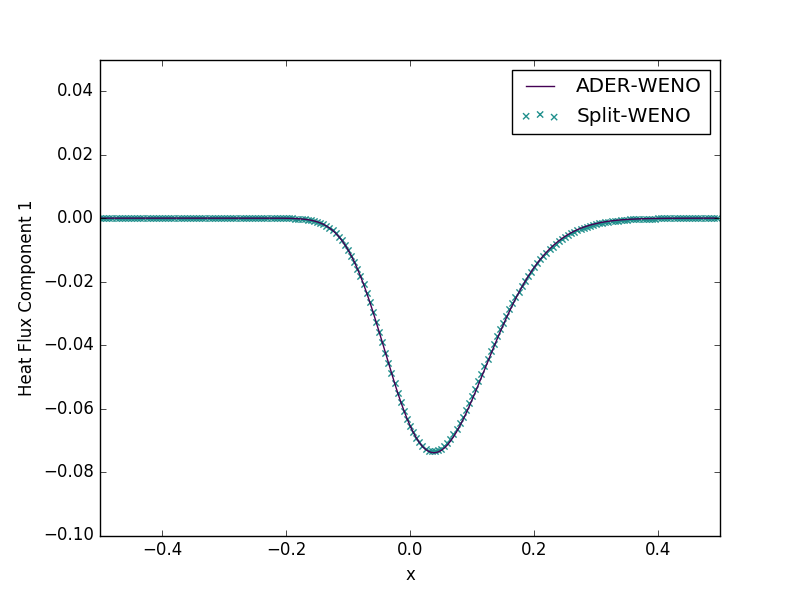}
\par\end{centering}
\caption{\label{fig:Heat-Conduction}Results of solving the problem of Heat
Conduction in Gas with both an ADER-WENO scheme and a Split-WENO scheme
($N=2$)}
\end{figure}

The material parameters are taken to be: $\gamma=1.4$, $c_{v}=2.5$,
$\rho_{0}=1$, $p_{0}=1$, $c_{s}=1$, $\alpha=2$, $\mu=10^{-2}$,
$\kappa=10^{-2}$ (resulting in $\tau_{1}=0.06$, $\tau_{2}=0.0025$).
The results of a simulation with 200 cells at time $t=1$, using reconstruction
polynomials of order $N=2$, are presented in \prettyref{fig:Heat-Conduction}.
The ADER-WENO and Split-WENO methods are in perfect agreement for
both the temperature and heat flux profiles. As demonstrated in \citet{Dumbser2016a},
this means that they in turn agree very well with a reference Navier-Stokes-Fourier
solution.

\subsection{Speed}

Both the ADER-WENO scheme and the Split-WENO scheme used in this study
were implemented in Python3. All array functions were precompiled
with Numba's JIT capabilities and the root-finding procedure in the
Galerkin predictor was performed using SciPy's Newton-Krylov solver,
compiled against the Intel MKL. Clear differences in computational
cost between the ADER-WENO and Split-WENO methods were apparent, as
is to be expected, owing to the lack of Galerkin method in the Split-WENO
scheme. The wall times for the various tests undertaken in this study
are given in \prettyref{tab:WallTimes}, comparing the combined WENO
and Galerkin methods of the ADER-WENO scheme to the combined WENO
and ODE methods of the Split-WENO scheme. All computations were performed
using an Intel Core i7-4910MQ, on a single core. The number of time
steps taken are given in \prettyref{tab:TimeSteps}. The differences
between the methods in terms of the number of time steps taken in
each test result from the fact that, for numerical stability, CFL
numbers of 0.8 and 0.7 were required by the ADER-WENO method and the
Split-WENO method, respectively.

Note that, unlike with the ADER-WENO scheme, the wall time for the
Split-WENO scheme is unaffected by a decrease in the viscosity in
Stokes' First Problem (and the corresponding increase in the stiffness
of the source terms). This is because the analytic approximation to
the distortion ODEs obviates the need for a stiff solver. The large
difference in ADER-WENO solver times between the $\mu=10^{-3}$ and
$\mu=10^{-4}$ cases is due to the fact that, in the latter case,
a stiff solver must be employed for the initial guess to the root
of the nonlinear system produced by the Discontinuous Galerkin method
(as described in \citet{Hidalgo2011}).

\begin{table}[p]
\begin{centering}
\bigskip{}
\begin{tabular}{|c|c|c|c|}
\hline 
 &
ADER-WENO &
Split-WENO &
Speed-up\tabularnewline
\hline 
\hline 
Stokes' First Problem ($\mu=10^{-2}$) &
265s &
38s &
7.0\tabularnewline
\hline 
Stokes' First Problem ($\mu=10^{-3}$) &
294s &
38s &
7.7\tabularnewline
\hline 
Stokes' First Problem ($\mu=10^{-4}$) &
536s &
38s &
14.1\tabularnewline
\hline 
Viscous Shock &
297s &
56s &
5.3\tabularnewline
\hline 
Heat Conduction in a Gas &
544s &
94s &
5.8\tabularnewline
\hline 
\end{tabular}
\par\end{centering}
\caption{\label{tab:WallTimes}Wall time for various tests (all with 200 cells)
under the ADER-WENO method and the Split-WENO method}
\medskip{}
\end{table}
\begin{table}[p]
\begin{centering}
\bigskip{}
\begin{tabular}{|c|c|c|}
\hline 
 &
Timesteps (ADER-WENO) &
Timesteps (Split-WENO)\tabularnewline
\hline 
\hline 
Stokes' First Problem ($\mu=10^{-2}$) &
385 &
442\tabularnewline
\hline 
Stokes' First Problem ($\mu=10^{-3}$) &
386 &
443\tabularnewline
\hline 
Stokes' First Problem ($\mu=10^{-4}$) &
385 &
442\tabularnewline
\hline 
Viscous Shock &
562 &
645\tabularnewline
\hline 
Heat Conduction in a Gas &
942 &
1077\tabularnewline
\hline 
\end{tabular}
\par\end{centering}
\caption{\label{tab:TimeSteps}Time steps taken for various tests (all with
200 cells) under the ADER-WENO method and the Split-WENO method}
\medskip{}
\end{table}

\subsection{Convergence\label{subsec:Convergence}}

To assess the rate of convergence of the Split-WENO method, the convected
isentropic vortex convergence study from \citet{Dumbser2016a} was
performed. The initial conditions are given as $\rho=1+\delta\rho$,
$p=1+\delta p$, $\boldsymbol{v}=\left(1,1,0\right)+\delta\boldsymbol{v}$,
$A=\sqrt[3]{\rho}I$, $\boldsymbol{J}=\boldsymbol{0}$, where:

\begin{subequations}

\begin{align}
\delta T & =-\frac{\left(\gamma-1\right)\epsilon^{2}}{8\gamma\pi^{2}}e^{1-r^{2}}\\
\delta\rho & =\left(1+\delta T\right)^{\frac{1}{\gamma-1}}-1\\
\delta p & =\left(1+\delta T\right)^{\frac{\gamma}{\gamma-1}}-1\\
\delta\boldsymbol{v} & =\frac{\epsilon}{2\pi}e^{\frac{1-r^{2}}{2}}\left(\begin{array}{c}
\begin{array}{c}
-\left(y-5\right)\\
x-5\\
0
\end{array}\end{array}\right)
\end{align}

\end{subequations}

The 2D domain is taken to be $\left[0,10\right]^{2}$. $\epsilon$
is taken to be 5. The material parameters are taken to be: $\gamma=1.4$,
$c_{v}=2.5$, $\rho_{0}=1$, $p_{0}=1$, $c_{s}=0.5$, $\alpha=1$,
$\mu=10^{-6}$, $\kappa=10^{-6}$ (resulting in $\tau_{1}=2.4\times10^{-5}$,
$\tau_{2}=10^{-6}$). Thus, this can be considered to be a stiff test
case.

The convergence rates in the $L_{1}$, $L_{2}$, $L_{\infty}$ norms
for the density variable are given in \prettyref{tab:ConvergenceN=00003D2}
and \prettyref{tab:ConvergenceN=00003D3} for WENO reconstruction
polynomial orders of $N=2$ and $N=3$, respectively. As expected,
both sets of tests attain roughly second order convergence. For comparison,
the corresponding results for this test from \citet{Dumbser2016a}
- solved using a third-order P2P2 scheme - are given in \prettyref{tab:ConvergenceN=00003D2-P2P2}
for comparison.

\begin{table}[p]
\begin{centering}
\bigskip{}
\begin{tabular}{|c|c|c|c|c|c|c|}
\hline 
Grid Size &
$\epsilon\left(L_{1}\right)$ &
$\epsilon\left(L_{2}\right)$ &
$\epsilon\left(L_{\infty}\right)$ &
$\mathcal{O}\left(L_{1}\right)$ &
$\mathcal{O}\left(L_{2}\right)$ &
$\mathcal{O}\left(L_{\infty}\right)$\tabularnewline
\hline 
\hline 
20 &
$2.87\times10^{-3}$ &
$7.15\times10^{-3}$ &
$6.21\times10^{-2}$ &
 &
 &
\tabularnewline
\hline 
40 &
$5.81\times10^{-4}$ &
$1.62\times10^{-3}$ &
$1.73\times10^{-2}$ &
2.30 &
2.14 &
1.85\tabularnewline
\hline 
60 &
$1.98\times10^{-4}$ &
$5.39\times10^{-4}$ &
$5.94\times10^{-3}$ &
2.65 &
2.70 &
2.63\tabularnewline
\hline 
80 &
$1.23\times10^{-4}$ &
$3.47\times10^{-4}$ &
$3.41\times10^{-3}$ &
1.67 &
1.52 &
1.92\tabularnewline
\hline 
\end{tabular}
\par\end{centering}
\caption{\label{tab:ConvergenceN=00003D2}Convergence rates for the Split-WENO
method ($N=2$)}
\medskip{}
\end{table}

\begin{table}[p]
\begin{centering}
\bigskip{}
\begin{tabular}{|c|c|c|c|c|c|c|}
\hline 
Grid Size &
$\epsilon\left(L_{1}\right)$ &
$\epsilon\left(L_{2}\right)$ &
$\epsilon\left(L_{\infty}\right)$ &
$\mathcal{O}\left(L_{1}\right)$ &
$\mathcal{O}\left(L_{2}\right)$ &
$\mathcal{O}\left(L_{\infty}\right)$\tabularnewline
\hline 
\hline 
10 &
$1.01\times10^{-2}$ &
$2.58\times10^{-2}$ &
$1.27\times10^{-1}$ &
 &
 &
\tabularnewline
\hline 
20 &
$1.68\times10^{-3}$ &
$4.02\times10^{-3}$ &
$2.93\times10^{-2}$ &
2.59 &
2.68 &
2.11\tabularnewline
\hline 
30 &
$5.34\times10^{-4}$ &
$1.57\times10^{-3}$ &
$1.70\times10^{-2}$ &
2.83 &
2.32 &
1.34\tabularnewline
\hline 
40 &
$3.32\times10^{-4}$ &
$8.94\times10^{-4}$ &
$7.55\times10^{-3}$ &
1.65 &
1.95 &
2.82\tabularnewline
\hline 
\end{tabular}
\par\end{centering}
\caption{\label{tab:ConvergenceN=00003D3}Convergence rates for the Split-WENO
method ($N=3$)}
\medskip{}
\end{table}

\begin{table}[p]
\begin{centering}
\bigskip{}
\begin{tabular}{|c|c|c|c|c|c|c|}
\hline 
Grid Size &
$\epsilon\left(L_{1}\right)$ &
$\epsilon\left(L_{2}\right)$ &
$\epsilon\left(L_{\infty}\right)$ &
$\mathcal{O}\left(L_{1}\right)$ &
$\mathcal{O}\left(L_{2}\right)$ &
$\mathcal{O}\left(L_{\infty}\right)$\tabularnewline
\hline 
\hline 
20 &
$9.44\times10^{-3}$ &
$2.20\times10^{-3}$ &
$2.16\times10^{-3}$ &
 &
 &
\tabularnewline
\hline 
40 &
$1.95\times10^{-3}$ &
$4.50\times10^{-4}$ &
$4.27\times10^{-4}$ &
2.27 &
2.29 &
2.34\tabularnewline
\hline 
60 &
$7.52\times10^{-4}$ &
$1.74\times10^{-4}$ &
$1.48\times10^{-4}$ &
2.35 &
2.35 &
2.61\tabularnewline
\hline 
80 &
$3.72\times10^{-4}$ &
$8.66\times10^{-5}$ &
$7.40\times10^{-5}$ &
2.45 &
2.42 &
2.41\tabularnewline
\hline 
\end{tabular}
\par\end{centering}
\caption{\label{tab:ConvergenceN=00003D2-P2P2}Convergence rates for the ADER-DG
PNPM method ($N,M=2$)}
\medskip{}
\end{table}

\section{Conclusions}

In summary, a new numerical method based on an operator splitting,
and including some analytical results, has been proposed for the GPR
model of continuum mechanics. It has been demonstrated that this method
is able to match current ADER-WENO methods in terms of accuracy on
a range of test cases. It is significantly faster than the other currently
available methods, and it is easier to implement. The author would
recommend that if very high order-of-accuracy is required, and computational
cost is not important, then ADER-WENO methods may present a better
option, as by design the new method cannot achieve better than second-order
accuracy. This new method clearly has applications in which it will
prove useful, however.

In a similar manner to the operator splitting method presented in
\citet{Leveque1990}, the Split-WENO method is second-order accurate
and stable even for very stiff problems (in particular, the reader
is referred to the results of the $\mu=10^{-4}$ variation of Stokes'
First Problem in \prettyref{subsec:Stokes'-First-Problem} and the
convergence study in \prettyref{subsec:Convergence}). However, it
will inevitably suffer from the incorrect speed of propagation of
discontinuities on regular, structured grids. This is due to a lack
of spatial resolution in evaluating the source terms, as detailed
in \citet{Leveque1990}. This issue can be rectified by the use of
some form of shock tracking or mesh refinement, as noted in the cited
paper. It is noted in \citet{Dumbser2008a} that operator splitting-based
methods can result in schemes that are neither well-balanced nor asymptotically
consistent. The extent to which these two conditions are violated
by the Split-WENO method - and the severity in practise of any potential
violation - is a topic of further research.

It should be noted that the assumption \eqref{eq:Assumption} used
to derive the approximate analytical solver may break down for situations
where the flow is compressed heavily in one direction but not the
others. The reason for this is that one of the singular values of
the distortion tensor will be much larger than the others, and the
mean of the squares of the singular values will not be close to its
geometric mean, meaning that the subsequent linearization of the ODE
governing the mean of the singular values fails. It should be noted
that none of the situations covered in this study presented problems
for the approximate analytical solver, and situations which may be
problematic are in some sense unusual. In any case, a stiff ODE solver
can be used to solve the system \eqref{eq:2odes} if necessary, utilizing
the Jacobians derived in the appendix, and so the Split-WENO method
is still very much usable in these situations, albeit slightly slower.

It should be noted that both the ADER-WENO and Split-WENO methods,
as described in this study, are trivially parallelizable on a cell-wise
basis. Thus, given a large number of computational cores, deficiencies
in the Split-WENO method in terms of its order of accuracy may be
overcome by utilizing a larger number of computational cells and cores.
The computational cost of each time step is significantly smaller
than with the ADER-WENO method, and the number of grid cells that
can be used scales roughly linearly with number of cores, at constant
time per iteration.

\section{References}

\bibliographystyle{elsarticle-harv}
\bibliography{23_home_hari_Git_papers_refs,24_home_hari_Git_papers_A_Fast_Numerical_Scheme___nski_Model_of_Continuum_Mechanics_misc_refs}

\begin{thebibliography}{28}
\expandafter\ifx\csname natexlab\endcsname\relax\def\natexlab#1{#1}\fi
\expandafter\ifx\csname url\endcsname\relax
  \def\url#1{\texttt{#1}}\fi
\expandafter\ifx\csname urlprefix\endcsname\relax\def\urlprefix{URL }\fi

\bibitem[{Balsara et~al.(2009)Balsara, Rumpf, Dumbser, and Munz}]{Balsara2009}
Balsara, D.~S., Rumpf, T., Dumbser, M., Munz, C.~D., 2009. {Efficient, high
  accuracy ADER-WENO schemes for hydrodynamics and divergence-free
  magnetohydrodynamics}. Journal of Computational Physics 228~(7), 2480--2516.
\newline\urlprefix\url{http://dx.doi.org/10.1016/j.jcp.2008.12.003}

\bibitem[{Barton and Drikakis(2010)}]{Barton2010}
Barton, P.~T., Drikakis, D., 2010. {An Eulerian method for multi-component
  problems in non-linear elasticity with sliding interfaces}. Journal of
  Computational Physics 229~(15), 5518--5540.
\newline\urlprefix\url{http://dx.doi.org/10.1016/j.jcp.2010.04.012}

\bibitem[{Becker(1929)}]{Beker1929}
Becker, R., 1929. {Impact Waves and Detonation}. Zeitschrift f{\"{u}}r Physik
  8, 381.

\bibitem[{Boscheri et~al.(2016)Boscheri, Dumbser, and
  Loub{\`{e}}re}]{Boscheri2016}
Boscheri, W., Dumbser, M., Loub{\`{e}}re, R., 2016. {Cell centered direct
  Arbitrary-Lagrangian-Eulerian ADER-WENO finite volume schemes for nonlinear
  hyperelasticity}. Computers and Fluids 134-135, 111--129.

\bibitem[{Dumbser et~al.(2008{\natexlab{a}})Dumbser, Balsara, Toro, and
  Munz}]{Dumbser2008}
Dumbser, M., Balsara, D.~S., Toro, E.~F., Munz, C.~D., 2008{\natexlab{a}}. {A
  unified framework for the construction of one-step finite volume and
  discontinuous Galerkin schemes on unstructured meshes}. Journal of
  Computational Physics 227~(18), 8209--8253.

\bibitem[{Dumbser et~al.(2008{\natexlab{b}})Dumbser, Enaux, and
  Toro}]{Dumbser2008a}
Dumbser, M., Enaux, C., Toro, E.~F., 2008{\natexlab{b}}. {Finite volume schemes
  of very high order of accuracy for stiff hyperbolic balance laws}. Journal of
  Computational Physics 227~(8), 3971--4001.

\bibitem[{Dumbser et~al.(2014)Dumbser, Hidalgo, and Zanotti}]{Dumbser2014}
Dumbser, M., Hidalgo, A., Zanotti, O., 2014. {High order space-time adaptive
  ADER-WENO finite volume schemes for non-conservative hyperbolic systems}.
  Computer Methods in Applied Mechanics and Engineering 268, 359--387.

\bibitem[{Dumbser et~al.(2016)Dumbser, Peshkov, Romenski, and
  Zanotti}]{Dumbser2016a}
Dumbser, M., Peshkov, I., Romenski, E., Zanotti, O., 2016. {High order ADER
  schemes for a unified first order hyperbolic formulation of continuum
  mechanics: Viscous heat-conducting fluids and elastic solids}. Journal of
  Computational Physics 314, 824--862.
\newline\urlprefix\url{http://arxiv.org/abs/1511.08995}

\bibitem[{Dumbser and Toro(2011{\natexlab{a}})}]{Dumbser2011a}
Dumbser, M., Toro, E.~F., 2011{\natexlab{a}}. {A simple extension of the Osher
  Riemann solver to non-conservative hyperbolic systems}. Journal of Scientific
  Computing 48~(1-3), 70--88.

\bibitem[{Dumbser and Toro(2011{\natexlab{b}})}]{Dumbser2011b}
Dumbser, M., Toro, E.~F., 2011{\natexlab{b}}. {On universal Osher-type schemes
  for general nonlinear hyperbolic conservation laws}. Communications in
  Computational Physics 10~(3), 635--671.

\bibitem[{Dumbser et~al.(2013)Dumbser, Zanotti, Hidalgo, and
  Balsara}]{Dumbser2013}
Dumbser, M., Zanotti, O., Hidalgo, A., Balsara, D.~S., 2013. {ADER-WENO finite
  volume schemes with space-time adaptive mesh refinement}. Journal of
  Computational Physics 248, 257--286.

\bibitem[{Frenkel(1947)}]{jacovfrenkel1947}
Frenkel, J., 1947. {Kinetic Theory of Liquids}. Oxford University Press.

\bibitem[{Giles(2008)}]{Giles2008}
Giles, M.~B., 2008. {An extended collection of matrix derivative results for
  forward and reverse mode algorithmic differentiation}. Tech. rep., University
  of Oxford.
\newline\urlprefix\url{http://eprints.maths.ox.ac.uk/1079/}

\bibitem[{Godunov and Romenski(2003)}]{godunov2003elements}
Godunov, S.~K., Romenski, E., 2003. {Elements of continuum mechanics and
  conservation laws}.

\bibitem[{Hidalgo and Dumbser(2011)}]{Hidalgo2011}
Hidalgo, A., Dumbser, M., 2011. {ADER schemes for nonlinear systems of stiff
  advection-diffusion-reaction equations}. Journal of Scientific Computing
  48~(1-3), 173--189.

\bibitem[{Jackson(2017)}]{Jackson2017}
Jackson, H., 2017. {On the Eigenvalues of the ADER-WENO Galerkin Predictor}.
  Journal of Computational Physics 333~(March 2017), 409--413.
\newline\urlprefix\url{http://dx.doi.org/10.1016/j.jcp.2016.12.058}

\bibitem[{Johnson(2013)}]{Johnson2013}
Johnson, B.~M., 2013. {Analytical shock solutions at large and small Prandtl
  number}. Journal of Fluid Mechanics 726, 1--12.
\newline\urlprefix\url{http://adsabs.harvard.edu/abs/2013arXiv1305.7132J}

\bibitem[{Leveque and Yee(1990)}]{Leveque1990}
Leveque, R., Yee, H., 1990. {A Study of Numerical Methods for Hyperbolic
  Conservation Laws with Stiff Source Terms}. Journal of Computational Physics
  86, 187--210.

\bibitem[{Malyshev and Romenskii(1986)}]{Malyshev1984}
Malyshev, A.~N., Romenskii, E.~I., 1986. {Hyperbolic equations for heat
  transfer. Global solvability of the Cauchy problem}. Siberian Mathematical
  Journal 27~(5), 734--740.

\bibitem[{McAdams et~al.(2011)McAdams, Selle, Tamstorf, Teran, and
  Sifakis}]{McAdams2011}
McAdams, A., Selle, A., Tamstorf, R., Teran, J., Sifakis, E., 2011. {Computing
  the Singular Value Decomposition of 3 x 3 matrices with minimal branching and
  elementary floating point operations}. University of Wisconsin Madison.

\bibitem[{Morduchow and Libby(1949)}]{Morduchow}
Morduchow, M., Libby, P.~A., 1949. {On a Complete Solution of the
  One-Dimensional Flow Equations of a Viscous, Heat-Conducting, Compressible
  Gas}. Tech. rep., Polytechnic Institute of Brooklyn.

\bibitem[{Oliphant(2007)}]{Jones}
Oliphant, T.~E., 2007. {SciPy: Open source scientific tools for Python}.
\newline\urlprefix\url{http://www.scipy.org/}

\bibitem[{Peshkov and Romenski(2016)}]{Peshkov2016}
Peshkov, I., Romenski, E., 2016. {A hyperbolic model for viscous Newtonian
  flows}. Continuum Mechanics and Thermodynamics 28~(1-2), 85--104.

\bibitem[{Romenski et~al.(2010)Romenski, Drikakis, and Toro}]{Romenski2010}
Romenski, E., Drikakis, D., Toro, E., 2010. {Conservative models and numerical
  methods for compressible two-phase flow}. Journal of Scientific Computing
  42~(1), 68--95.

\bibitem[{Romenski et~al.(2007)Romenski, Resnyansky, and Toro}]{Romenski2007}
Romenski, E., Resnyansky, A.~D., Toro, E.~F., 2007. {Conservative hyperbolic
  model for compressible two-phase flow with different phase pressures and
  temperatures}. Quarterly of applied mathematics 65(2)~(2), 259--279.

\bibitem[{Romenski(1989)}]{Romenski1988}
Romenski, E.~I., 1989. {Hyperbolic equations of Maxwell's nonlinear model of
  elastoplastic heat-conducting media}. Siberian Mathematical Journal 30~(4),
  606--625.

\bibitem[{Toro(2009)}]{Toro2009}
Toro, E., 2009. {Riemann solvers and numerical methods for fluid dynamics: a
  practical introduction}. Springer.

\bibitem[{Zanotti and Dumbser(2016)}]{Zanotti2016}
Zanotti, O., Dumbser, M., 2016. {Efficient conservative ADER schemes based on
  WENO reconstruction and space-time predictor in primitive variables}.
  Computational Astrophysics and Cosmology 3~(1), 1.
\newline\urlprefix\url{http://www.comp-astrophys-cosmol.com/content/3/1/1}

\end{thebibliography}

\section{Acknowledgments}

I acknowledge financial support from the EPSRC Centre for Doctoral
Training in Computational Methods for Materials Science under grant
EP/L015552/1.

\section{Appendix}

\subsection{Jacobian of Distortion ODEs\label{subsec:Jacobian-of-Distortion-ODEs}}

The Jacobian of the source function is used to speed up numerical
integration of the ODE. It is derived thus:

\begin{equation}
\frac{\partial\dev\left(G\right)_{ij}}{\partial A_{mn}}=\delta_{in}A_{mj}+\delta_{jn}A_{mi}-\frac{2}{3}\delta_{ij}A_{mn}
\end{equation}

Thus:

{\small{}
\begin{align}
\frac{\partial\left(A\dev\left(G\right)\right)_{ij}}{\partial A_{mn}} & =\frac{\partial A_{it}}{\partial A_{mn}}\dev\left(G\right)_{tj}+A_{it}\frac{\partial\dev\left(G\right)_{tj}}{\partial A_{mn}}\\
 & =\delta_{im}\delta_{tn}\left(A_{kt}A_{kj}-\frac{1}{3}A_{kl}A_{kl}\delta_{tj}\right)+A_{it}\left(\delta_{tn}A_{mj}+\delta_{jn}A_{mt}-\frac{2}{3}\delta_{tj}A_{mn}\right)\nonumber \\
 & =\delta_{im}A_{kn}A_{kj}-\frac{1}{3}\delta_{im}\delta_{jn}A_{kl}A_{kl}+A_{in}A_{mj}+\delta_{jn}A_{ik}A_{mk}-\frac{2}{3}A_{ij}A_{mn}\nonumber 
\end{align}
}{\small \par}

Thus:

\begin{align}
J_{A} & \equiv\frac{-3}{\tau_{1}}\frac{\partial\left(\det\left(A\right)^{\frac{5}{3}}A\dev\left(G\right)\right)_{ij}}{\partial A_{mn}}\\
 & =\frac{-3}{\tau_{1}}\det\left(A\right)^{\frac{5}{3}}\left(\frac{5}{3}\left(A\dev\left(G\right)\right)_{ij}A_{mn}^{-T}+A_{in}A_{mj}+\delta_{jn}G_{im}^{'}+\delta_{im}G_{jn}-\frac{1}{3}\delta_{im}\delta_{jn}A_{kl}A_{kl}-\frac{2}{3}A_{ij}A_{mn}\right)\nonumber \\
 & =\frac{1}{\tau_{1}}\det\left(A\right)^{\frac{5}{3}}\left(-5\left(A\dev\left(G\right)\right)\otimes A^{-T}+2A\otimes A-3\left(A\otimes A\right)^{1,3}+\left\Vert A\right\Vert _{F}^{2}\left(I\otimes I\right)^{2,3}-3\left(G^{'}\otimes I+I\otimes G\right)^{2,3}\right)\nonumber 
\end{align}

where $G^{'}=AA^{T}$ and $X^{a,b}$ refers to tensor $X$ with indices
$a,b$ transposed.

\subsection{Jacobian of Thermal Impulse ODEs\label{subsec:Jacobian-of-Thermal-Impulse-ODEs}}

As demonstrated in \prettyref{subsec:The-Thermal-Impulse-ODEs}, we
have:

\begin{equation}
\frac{dJ_{i}}{dt}=\frac{J_{i}}{2}\left(-a+b\left(J_{1}^{2}+J_{2}^{2}+J_{3}^{2}\right)\right)
\end{equation}

where

\begin{subequations}

\begin{align}
a & =\frac{2\rho_{0}}{\tau_{2}T_{0}\rho c_{v}}\left(E-E_{2A}\left(A\right)-E_{3}\left(\boldsymbol{v}\right)\right)\\
b & =\frac{\rho_{0}\alpha^{2}}{\tau_{2}T_{0}\rho c_{v}}
\end{align}

\end{subequations}

Thus, the Jacobian of the thermal impulse ODEs is:

\begin{equation}
\left(\begin{array}{ccc}
\frac{b}{2}\left(3J_{1}^{2}+J_{2}^{2}+J_{3}^{2}\right)-\frac{a}{2} & bJ_{1}J_{2} & bJ_{1}J_{3}\\
bJ_{1}J_{2} & \frac{b}{2}\left(J_{1}^{2}+3J_{2}^{2}+J_{3}^{2}\right)-\frac{a}{2} & bJ_{2}J_{3}\\
bJ_{1}J_{3} & bJ_{2}J_{3} & \frac{b}{2}\left(J_{1}^{2}+J_{2}^{2}+3J_{3}^{2}\right)-\frac{a}{2}
\end{array}\right)
\end{equation}

\subsection{WENO Matrices for $N=2$\label{subsec:WENO-Matrices-for-N=00003D2}}

\begin{subequations}

\begin{align}
M_{1} & =\left(\begin{array}{ccc}
2\sqrt{\frac{5}{3}}+\frac{245}{18} & -\frac{236}{9} & \frac{245}{18}-2\sqrt{\frac{5}{3}}\\
\sqrt{\frac{5}{3}}+\frac{65}{18} & -\frac{56}{9} & \frac{65}{18}-\sqrt{\frac{5}{3}}\\
\frac{5}{18} & \frac{4}{9} & \frac{5}{18}
\end{array}\right)\\
M_{2} & =\left(\begin{array}{ccc}
\sqrt{\frac{5}{3}}+\frac{65}{18} & -\frac{56}{9} & \frac{65}{18}-\sqrt{\frac{5}{3}}\\
\frac{5}{18} & \frac{4}{9} & \frac{5}{18}\\
\frac{65}{18}-\sqrt{\frac{5}{3}} & -\frac{56}{9} & \sqrt{\frac{5}{3}}+\frac{65}{18}
\end{array}\right)\\
M_{3} & =\left(\begin{array}{ccc}
\frac{5}{18} & \frac{4}{9} & \frac{5}{18}\\
\frac{65}{18}-\sqrt{\frac{5}{3}} & -\frac{56}{9} & \sqrt{\frac{5}{3}}+\frac{65}{18}\\
\frac{245}{18}-2\sqrt{\frac{5}{3}} & -\frac{236}{9} & 2\sqrt{\frac{5}{3}}+\frac{245}{18}
\end{array}\right)
\end{align}

\end{subequations}

\begin{subequations}

\begin{align}
M_{1}^{-1} & =\left(\begin{array}{ccc}
\frac{1}{60}\left(2-3\sqrt{15}\right) & \sqrt{\frac{3}{5}}-\frac{1}{15} & \frac{1}{60}\left(62-9\sqrt{15}\right)\\
-\frac{1}{24} & \frac{1}{12} & \frac{23}{24}\\
\frac{1}{60}\left(3\sqrt{15}+2\right) & -\sqrt{\frac{3}{5}}-\frac{1}{15} & \frac{1}{60}\left(9\sqrt{15}+62\right)
\end{array}\right)\\
M_{2}^{-1} & =\left(\begin{array}{ccc}
\frac{1}{60}\left(3\sqrt{15}+2\right) & \frac{14}{15} & \frac{1}{60}\left(2-3\sqrt{15}\right)\\
-\frac{1}{24} & \frac{13}{12} & -\frac{1}{24}\\
\frac{1}{60}\left(2-3\sqrt{15}\right) & \frac{14}{15} & \frac{1}{60}\left(3\sqrt{15}+2\right)
\end{array}\right)\\
M_{3}^{-1} & =\left(\begin{array}{ccc}
\frac{1}{60}\left(9\sqrt{15}+62\right) & -\sqrt{\frac{3}{5}}-\frac{1}{15} & \frac{1}{60}\left(3\sqrt{15}+2\right)\\
\frac{23}{24} & \frac{1}{12} & -\frac{1}{24}\\
\frac{1}{60}\left(62-9\sqrt{15}\right) & \sqrt{\frac{3}{5}}-\frac{1}{15} & \frac{1}{60}\left(2-3\sqrt{15}\right)
\end{array}\right)
\end{align}

\end{subequations}
\end{document}